\documentclass[a4paper,twocolumn,11pt,accepted=2025-07-11]{quantumarticle}
\pdfoutput=1
\usepackage[utf8]{inputenc}
\usepackage[english]{babel}
\usepackage[T1]{fontenc}
\usepackage{amsmath}
\usepackage{hyperref}
\usepackage{tikz}
\usepackage{lipsum}
\usepackage{threeparttable}
\usepackage{bm}
\usepackage{braket}
\usepackage{multirow}
\usepackage{bigdelim}
\usepackage{subfigure}
\usepackage{comment}
\usepackage{here}
\usepackage{cite}

\begin{document}

\title{Quick design of feasible tensor networks for constrained combinatorial optimization}

\author{Hyakka Nakada}
\affiliation{Recruit Co., Ltd., Tokyo 100-6640, Japan}
\affiliation{Graduate School of Science and Technology, Keio University, Kanagawa 223-8522, Japan}
\email{hyakka\_nakada@r.recruit.co.jp}
\author{Kotaro Tanahashi}
\affiliation{Recruit Co., Ltd., Tokyo 100-6640, Japan}
\author{Shu Tanaka}
\affiliation{Graduate School of Science and Technology, Keio University, Kanagawa 223-8522, Japan}
\affiliation{Department of Applied Physics and Physico-Informatics, Keio University, Kanagawa 223-8522, Japan}
\affiliation{Keio University Sustainable Quantum Artificial Intelligence Center (KSQAIC), Keio University, Tokyo 108-8345, Japan}
\affiliation{Human Biology-Microbiome-Quantum Research Center (WPI-Bio2Q), Keio University, Tokyo 108-8345, Japan}

\maketitle

\begin{abstract}
Quantum computers are expected to enable fast solving of large-scale combinatorial optimization problems.
However, their limitations in fidelity and the number of qubits prevent them from handling real-world problems.
Recently, a quantum-inspired solver using tensor networks has been proposed, which works on classical computers.
Particularly, tensor networks have been applied to constrained combinatorial optimization problems for practical applications. By preparing a specific tensor network to sample states that satisfy constraints, feasible solutions can be searched for without the method of penalty functions. 
Previous studies have been based on profound physics, such as U(1) gauge schemes and high-dimensional lattice models.
In this study, we devise to design feasible tensor networks using elementary mathematics without such a specific knowledge.
One approach is to construct tensor networks with nilpotent-matrix manipulation. The second is to algebraically determine tensor parameters. 
We showed mathematically that such feasible tensor networks can be constructed to accommodate various types of constraints.
For the principle verification, we numerically constructed a feasible tensor network for facility location problem, to find much faster construction than conventional methods.
Then, by performing imaginary time evolution, feasible solutions were always obtained, ultimately leading to the optimal solution. 
\end{abstract}

\section{Introduction}
\label{sec:sec1}
Combinatorial optimization is the process of identifying a set of discrete variables that minimizes or maximizes an objective function. Numerous real-world problems can be viewed as combinatorial optimization, which has a significant academic and industrial importance. In modern society, as the amount of data traffic increases due to technological advances, so does the size of the combinatorial optimization problems to be solved.
Therefore, fast optimization solvers are widely researched, such as integer programming and metaheuristics. 
In recent years, quantum-gate-type computers have been attracting much attention and are expected to solve combinatorial optimization problems on such a large scale that they cannot be handled by classical computers. 

However, the current quantum-gate hardware has a small number of quantum bits and errors during execution. Such hardware is called Noisy Intermediate-Scale Quantum (NISQ) devices~\cite{Preskil2018nisq} and solving practical problems remains challenging.
This leads to motivating many quantum-inspired classical algorithms to appear for baseline, approximate, and alternative methodologies in classical computers.
For example, methods using Tensor Networks (TNs)~\cite{bridgeman2017tn, orus2019tn, okunishi2022tn} have been proposed. TNs can approximate quantum simulations by limiting the coefficients of quantum states into the form of tensor products and utilizing singular value decomposition. Thus, it is possible to solve combinatorial optimization problems at a scale beyond the capabilities of current NISQ devices. 

Recently, TN based solvers specialized for combinatorial optimization have been actively researched~\cite{minato2024tnsolver, yasuda2024tnsolver, morais2025tnsolver}.
Other efforts have been reported, including approaches such as searching for ground states by differentiable programming~\cite{liu2021tropical} and classically simulating Quantum Approximate Optimization Algorithm (QAOA)~\cite{lykov2022tnqaoa}.
Particularly, in the weighted Max-Cut problem, a TN approach has been reported to consistently yield high approximation ratios and efficient execution on larger graphs, compared with other classical solvers~\cite{morais2025tnsolver}. 
Not only to mimic quantum-gate methods, but also to find better optimization algorithms, TNs are gaining much attention as a new solver for combinatorial optimization problems.

Many real-world problems have constraints and require solutions within the feasible solution space that satisfies these constraints.
In the field of quantum or quantum-inspired algorithms, constrained combinatorial optimization is typically solved by the penalty function method~\cite{bertsekas1982penalty, luenberger2008penalty, lucas2014penalty, tanaka2017penalty, takehara2019penalty, tamura2021penalty, tanahashi2019penalty, zaman2022penalty}. In this method, violation terms for constraint conditions are added to the original cost Hamiltonian, allowing feasible solutions to be effectively searched for. 
However, several challenges arise such as difficulty in adjusting the penalty coefficients, an increase in computational cost due to interactions between many qubits, and inability to completely prohibit infeasible solutions. 

To address these challenges, a lot of algorithms without the penalty function method have been proposed for quantum algorithms~\cite{hadfield2019qaoa, wang2020qaoa, bartschi2020grover, matsuo2020pqc, nakada2025pqc} and quantum-inspired algorithms~\cite{hao2022tnopt, lopez2023tnopt, lopez2024tnopt, bachmayr2022particle}.
One such method involves preparing TNs that describe the superpositioned state of feasible solutions and searching for the ground state with imaginary time evolution~\cite{hao2022tnopt}. 
Under ideal imaginary time evolution, transitions to the states of infeasible solutions are forbidden.
The second method employs a similar TN as a generative model and iteratively modifies the parameters of each tensor based on the energy expectation value~\cite{lopez2023tnopt}. This approach is an application of the Generator-Enhanced Optimization (GEO)~\cite{alcazar2022geo} to constrained combinatorial optimization problems. Both methods efficiently optimize by structuring TNs to exclusively output states of feasible solutions. In other words, the capability to design a ``feasible'' TN is crucial for solving constrained combinatorial optimization problems. 

In addition to special constraints such as cardinality ones~\cite{lopez2023tnopt, bachmayr2022particle}, other constraints have been targeted. For example, TNs are applied to open-pit mining problem by introducing additional tensors that represent flags to indicate whether the constraint is satisfied or not~\cite{hao2022tnopt}. Another method has been reported to handle arbitrary linear constraints as well as cardinality ones by introducing U(1) gauge symmetry, which ensures the law of particle number conservation~\cite{lopez2023tnopt}.
However, the first method requires an auxiliary tensor to connect the physical variables appearing in the constraints. 
This leads to an exponential increase in their tensor sizes in the case of global constraints, although the open-pit mining problem has only local constraints. Furthermore, the TNs have a higher dimensional structure than Matrix Product State (MPS).
Although the second method can encode global linear constraints to an efficient MPS, it requires processing such as backtracking to ensure U(1) gauge symmetry. This causes deriving ``feasible'' TNs is $\sharp$P-hard when multiple constraints are imposed~\cite{lopez2023tnopt}.

\begin{figure*}[t]
\centering
\includegraphics[width=14cm]{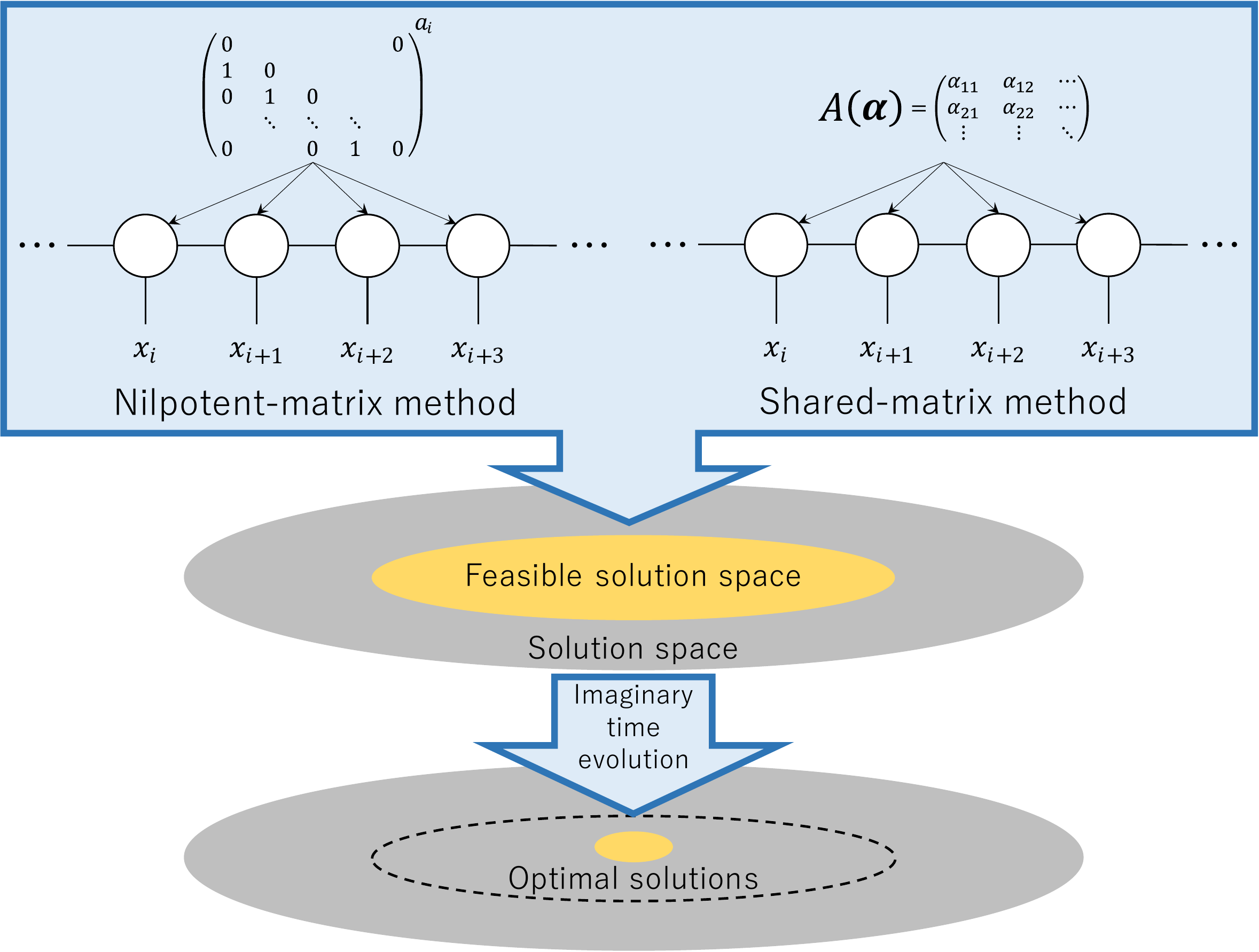}
\caption{Overview of proposed method. Feasible solutions can be encoded by Tensor Networks (TNs) through nilpotent-matrix method and shared-matrix method. Then, imaginary time evolution is applied to TNs so as to find the optimal solutions. Optimization methods are not limited to imaginary time evolution.}
\label{fig:figure1}
\end{figure*}
As mentioned above, the previous researches applying TNs to constrained combinatorial optimization mainly originate from physics schemes. In this study, we propose designing ``feasible'' TNs using only elementary mathematics, which leads to more user-friendly design of TNs and application to a wider range of constraints.
Thanks to the user-friendly design, the extensibility of constraint conditions can be improved. To deal with real-world problems, tasks are often designed through trial and error, such as the addition of another new constraint, because the appropriate constraint conditions of the problem are usually not predetermined. 
However, in the previous methods~\cite{hao2022tnopt, lopez2023tnopt}, it is necessary to redesign the fully ``feasible'' TNs from scratch to satisfy all constraints. Because the proposed method can quickly obtain them by simple mathematical manipulation, further utilization of TNs is expected in the field of combinatorial optimization.

Figure~\ref{fig:figure1} provides an overview of the proposed methods. The first is a method that adopts a nilpotent matrix as a tensor. Specifically, we employ a nilpotent matrix to the power of the coefficient $a_i$ in a constraint equation. Thus, ``feasible'' TNs can be obtained without any backtracking algorithms. This method is a modification of the conventional TNs conserving the number of particles, which enables the handling of a linear inequality or equality constraint. To consider the presence of more complex constraints beyond a single linear constraint, as a second method, we have devised an approach that adopts shared matrices for the tensors of each site. Their parameters $\bm{\alpha}$ are determined algebraically, without relying on the physics scheme of particle number conservation. In other words, this is a method to formulate and solve equations for parameters so that the coefficients of the states that satisfy the constraint are non-zero, while others are set to $0$. 

After preparing ``feasible'' TNs, optimization is performed.
Here, the optimization method is not limited to imaginary time evolution. Alternatively, one can use such as density matrix renormalization group~\cite{bridgeman2017tn} or GEO method~\cite{alcazar2022geo}.

In this study, our purpose is creating ``feasible'' TNs for wider kinds of constraints and demonstrating optimization for the principle verification.
By the above methods, we showed mathematically that ``feasible'' TNs can be constructed to accommodate various types of constraints, such as linear equality and inequality constraints, comparison constraints regarding the magnitude relationship between variables, and so on. 
We numerically constructed a ``feasible'' TN for facility location problem, to find much faster construction than conventional methods.
Then, by applying imaginary time evolution to the obtained TN, the optimal solutions can be sought.
Feasible solutions were consistently obtained, and the optimal solutions were achieved after sufficient time evolution.

The proposed method can construct a ``feasible'' TN without the penalty function method. Moreover, the results of this study may not only contribute to the development of optimization methods using TNs but also lead to that using quantum gates. In recent years, a technique for converting TNs into equivalent quantum circuits has been proposed~\cite{rudolph2023tnqc}. By combining such a technique and the proposed method, we can devise a new solver for constrained combinatorial optimization using quantum gates.

We describe the structure of this paper. First, in Section~\ref{sec:sec2}, we introduce the definition of ``feasible'' TNs and the previous studies. In Section~\ref{sec:sec3} and \ref{sec:sec4}, we discuss the theory of constructing feasible TNs using nilpotent-matrix method and shared-matrix method, respectively. 
Then, in Section~\ref{sec:sec5}, we extend to multiple constraints.
Section~\ref{sec:sec6} details the costs of TNs and compare them with baselines. 
In Section~\ref{sec:sec7}, we solve facility location problem and explain the results while comparing with the baselines. Finally, we conclude in Section~\ref{sec:sec8}.

\section{Preliminaries}
\label{sec:sec2}
\subsection{Tensor network}
\label{sec:subsec2.1}

TNs are mathematical expressions used to describe entangled quantum states in many-body systems~\cite{bridgeman2017tn, orus2019tn, okunishi2022tn}. Well-known TNs include MPS and Pair Entangled Projected State (PEPS)~\cite{verstraete2004peps}. A quantum state can generally be represented by
\begin{equation}
\label{eq:eq1}
\left| \psi\right\rangle=\sum_{\bm{x}}{\psi_{x_1,x_2,\ldots,x_N}\left|x_1,x_2,\ldots,x_N\right\rangle}.
\end{equation}
Here, each $x_i$ represents a physical variable that takes a bit value of $\{0,1\}$. 
Thus, summation over $\bm{x}$ is performed in $\{0,1\}^N$.
$\psi_{x_1,x_2,\ldots,x_N}$ represents the state coefficient. In TNs, this coefficient is restricted to the form of a product of tensors. The original quantum state shown in Eq.~\eqref{eq:eq1} requires memory of the order of $2^N$. By reducing the size of the tensor, quantum states can be approximately simulated with less memory.

In an MPS, the state coefficient in Eq.~\eqref{eq:eq1} is modeled by 
\begin{equation}
\label{eq:eq2}
\psi_{x_1,x_2,\ldots,x_N} = \operatorname{tr}\left[\prod_{i=1}^{N} A^{\left[i\right]x_i}\ \right].
\end{equation}
Here, $i$ represents the coordinates of each site in the one-dimensional lattice system and the left and right bonds of on-site matrices $A^{\left[i\right]x_i}$ must be determined so that the trace is well-defined. 

Another example of a TN is the PEPS below.
\begin{equation}
\left| \psi\right\rangle=\sum_{\bm{x}} {\operatorname{tr}' \left[\prod_{i=1}^{N_1}\prod_{j=1}^{N_2} A^{[i,j] x_{i,j}} \right] \left|x_{1,1},\ldots,x_{N_1,N_2}\right\rangle} . \nonumber
\end{equation}
Here, $i,j$ represents the coordinates of each site in the two-dimensional lattice system. Although the MPS uses a regular trace, in the case of PEPS, a more general trace $\operatorname{tr}'$ is used. This tensor trace contracts dummy variables not only in the left and right bonds but also in the up and down bonds.
The bonds of on-site matrices $A^{\left[i,j\right]x_{i,j}}$ must be determined so that the trace is well-defined. 

\subsection{Feasible TN for constrained combinatorial optimization}
\label{sec:subsec2.2}
We consider a TN where at least one coefficient of a feasible solution state for given constraint $C$ is a non-zero real value, and any coefficient of a state that violates the constraint is always $0$. This TN is defined as a ``feasible'' TN for the constraint $C$. 

Additionally, the special feasible TNs where the coefficients of any feasible solution state are non-zero are called ``fully feasible'' TNs. 
For finding the optimal solutions with imaginary time evolution, an initial TN must be fully feasible.
Thus, this paper focuses on constructing a fully feasible MPS. That is, in Eq.~\eqref{eq:eq2}, we aim to find an MPS that satisfies
\begin{equation}
\label{eq:eq3}
\operatorname{tr}\left[\prod_{i=1}^{N} A^{\left[i\right]x_i}\right] =
\begin{cases}  0 & \text{for $\forall \bm{x} \notin X_C$} , \\
\text{Non-zero} & \text{for $\forall \bm{x} \in X_C$} ,
\end{cases}
\end{equation}
where $X_C$ represents the set of all the feasible solutions that satisfy the constraint $C$.

\subsection{Previous studies on feasible TN construction}
\label{sec:subsec2.3}
As previous studies on designing feasible TNs, two main methods are explained. The first method constructs an MPS with $U(1)$ gauge symmetry to conserve the number of particles~\cite{lopez2023tnopt}. This ensures that the charge $N_{\mathrm{in}}$ coming in each site, the charge $N_{\mathrm{out}}$ going out, and the on-site charge $n$ are totally balanced. Such an MPS has the following forms, 
\begin{equation}
\label{eq:eq4}
\begin{split}
A_{\alpha,\beta}^a&=\left(A_{n_\alpha,n_\beta}^{n_a}\right)_{t_{n_\alpha},t_{n_\beta}}^{t_{n_a}}\delta_{n+N_{\mathrm{in}},N_{\mathrm{out}}}, \\
&N_{\mathrm{in}}=\sum_{i\in \mathcal{I} } n_i,N_{\mathrm{out}}=\sum_{i\in \mathcal{O} } n_i,
\end{split}
\end{equation}
where $t_n=1,2,\ldots,d_n$ is the index of degeneracy, and $d_n$ denotes the degree of degeneracy of the charge $n$. $\mathcal{I},\mathcal{O}$ are the subscripts of the sites coming in and going out, respectively.
$\alpha$ and $\beta$ are dummy variables.
In the case of a constant sum constraint, in other words cardinality constraint $\sum_{i=1}^N x_i=d$, 

\begin{equation}
\label{eq:eq5}
\scalebox{0.8}{$
\begin{split}
A^{[i]0}=
\begin{cases}  \left(\begin{matrix}0&1&&&\text{\huge{0}}\\&0&1&&\\&&\ddots&\ddots&\\\text{\huge{0}}&&&0&1\\\end{matrix}\right) & \text{for $i=1,\ldots,d$} , \\
\left(\begin{matrix}1&&&&\text{\huge{0}}\\&1&&&\\&&1&&\\&&&\ddots&\\\text{\huge{0}}&&&&1\\\end{matrix}\right) & \text{for $i=d+1,\ldots,N-d$} , \\
\left(\begin{matrix}0&&&\text{\huge{0}}\\1&0&&\\&\ddots&\ddots&\\&&1&0\\\text{\huge{0}}&&&1\\\end{matrix}\right)  & \text{for $i=N-d+1,\ldots,N$}
\end{cases} \\
A^{[i]1}=
\begin{cases}  \left(\begin{matrix}1&0&&&\text{\huge{0}}\\&1&0&&\\&&\ddots&\ddots&\\\text{\huge{0}}&&&1&0\\\end{matrix}\right) & \text{for $i=1,\ldots,d$} , \\
\left(\begin{matrix}0&&&&\text{\huge{0}}\\1&0&&&\\&1&0&&\\&&\ddots&\ddots&\\\text{\huge{0}}&&&1&0\\\end{matrix}\right) & \text{for $i=d+1,\ldots,N-d$} , \\
\left(\begin{matrix}1&&&\text{\huge{0}}\\0&1&&\\&\ddots&\ddots&\\&&0&1\\\text{\huge{0}}&&&0\\\end{matrix}\right)& \text{for $i=N-d+1,\ldots,N$}
\end{cases}
\end{split}
$}
\end{equation}
are one of the feasible tensors so as to satisfy Eq.~\eqref{eq:eq4}. The size of each matrix is $i\times (i+1)$ for $i=1,\ldots,d$, $d\times d$ for $i=d+1,\ldots,N-d$, and $(N-i+2)\times (N-i+1)$ for $i=N-d+1,\ldots,N$.
Here, each element value was conveniently set to $1$, but any real number is acceptable. 

While several efforts have reported to connect between linear constraints and TNs~\cite{biamonte2015linear, kourtis2019linear, ryzhakov2022linear, liu2023linear}, this method can be applied to arbitrary linear constraints.
However, it is necessary to find any consistent set of all link charges in the constraints. Finding the appropriate TN becomes $\sharp$P-hard. Therefore, the paper~\cite{lopez2023tnopt} stated that it is difficult to derive a fully feasible TN for multiple linear constraints.

Recently, an improved method has been reported to efficiently design fully feasible TNs without explicitly deriving all link charges~\cite{lopez2024tnopt}. However, as in the literature~\cite{lopez2023tnopt}, backtracking is used, which causes exponential computation for the number of the constraints. Thus, obtaining TNs is difficult when a large number of constraints are imposed. Handling general constraints other than linear ones is also difficult.
In addition, there is a weakness in constraint extensibility. When another new constraint is added to existing constraints, it should be noted that redesigning fully feasible TNs from scratch is necessary to satisfy all constraints.

As another previous study, a TN for constraints among local sites has also been proposed~\cite{hao2022tnopt}. In this method, auxiliary tensors that represent whether the constraints are satisfied or not are added. For example, the constraint $C:x_1+x_2+x_3=1$ is considered.
\begin{equation}
\label{eq:eq6}
\left | \psi\right\rangle=\sum_{\bm{x},\bm{y}} \prod_{i=1}^{3} A_{y_i}^{[i] x_i} B_{y_1,y_2,y_3}^{[C]} \left | x_1,x_2,x_3 \right\rangle .
\end{equation}
Both the physical variable $x_1,x_2,x_3$ and the dummy variable $y_1,y_2,y_3$ are binary variables of $\{0,1\}$. $B^{[C]}$ corresponds to the auxiliary tensor for the constraint condition $C$ and is responsible for connecting the physical variables that are involved in this condition. 
All the elements of tensors are initially set to $0$. Then, for the fully feasibleness,
$B_{0,0,1}^{[C]}=B_{0,1,0}^{[C]}=B_{1,0,0}^{[C]}=1$ and $A_0^{[1] 0}=A_1^{[1] 1}=A_0^{[2] 0}=A_1^{[2] 1}=A_0^{[3] 0}=A_1^{[3] 1}=1$ are encoded.
In the paper~\cite{hao2022tnopt}, a PEPS for the open-pit mining problem is constructed using such tensor structure. Several researches have been reported to apply this method to other problems~\cite{ali2024tnopt, ali2023tnopt}.

Theoretically, introducing auxiliary tensors can handle a general type of constraints. However, the bond size of $B^{[C]}$ becomes exponentially large for the number of physical variables involved in the constraint $C$. Thus, this method is considered to be specialized for local constraints.
When another new constraint is added to existing constraints, redesigning fully feasible TNs from scratch is necessary as with the methods in~\cite{lopez2023tnopt, lopez2024tnopt}.
In addition, introducing auxiliary tensors disables the computational advantages of MPS because higher dimensional lattices are required.

\section{Feasible MPS construction by nilpotent-matrix method}
\label{sec:sec3}
\begin{figure}[ht]
\centering
\includegraphics[width=8cm]{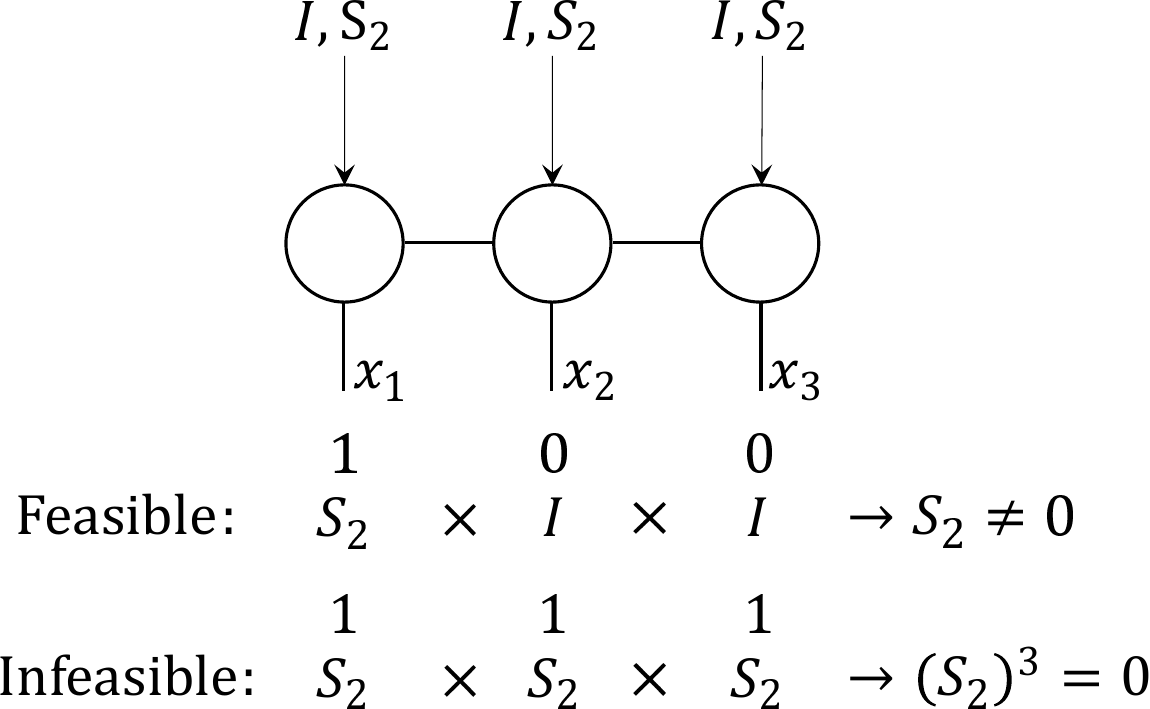}
\caption{MPS constructed by nilpotent-matrix method (equally and positively weighted). A special constraint $x_1+x_2+x_3 \leq 2$ is considered. 
A nilpotent matrix $S_2$ becomes zero when multiplied more times than upper bound $2$. 
}
\label{Nilpotent_matrix_MPS}
\label{fig:figure2}
\end{figure}
We explain how to construct a fully feasible MPS using the nilpotent-matrix method. Target constraints are linear inequality or equality constraints with integer coefficients.
Letting $d$ be an integer greater than $0$, one nilpotent matrix of exponent $d+1$ can be expressed by
\begin{equation}
\label{eq:eq7}
S_d \equiv \left(\begin{matrix}0&&&&\text{\huge{0}}\\1&0&&&\\0&1&0&&\\&\ddots&\ddots&\ddots&\\\text{\huge{0}}&&0&1&0\\\end{matrix}\right).
\end{equation}
This is a $(d+1) \times (d+1)$ matrix that becomes zero matrix when raised to the power of $d+1$. Although we set $1$ as a non-zero element, any real number can be set generally. In particular, if we set $\sqrt{1},\sqrt{2},\sqrt{3},\ldots$ in order from the top left of the matrix, it will be equivalent to the matrix representation of creation or annihilation operators. From Eq.~\eqref{eq:eq7}, the nilpotent matrix $S_d$ to the power of $k$ is
\begin{equation}
\label{eq:eq8}
\scalebox{0.95}{$
(S_d)^k = 
\begin{array}{ccccccccc}
 &\ldelim({7}{5pt}[] & & & & & &\text{\LARGE{0}} &\rdelim){7}{0pt}[] \\
 &  & & & & & & & \\
& &0 & & & & & & \\
\ldelim\{{4}{45pt}[$d-k+1$]& &1 &0 & & & & & \\
& &0 &1 &0 & & & & \\
& & &\ddots &\ddots &\ddots & & & \\
&  &\text{\LARGE{0}} & &0 &1 &0 & & 
\end{array}
$}
\end{equation}
where $k=0,1,\ldots,d$. Thus, $S_d$ has the property that the sub-diagonal elements with the value of $1$ descend one step towards the lower left, each time the matrix is multiplied.
As shown in Fig.~\ref{fig:figure2}, the nilpotent-matrix method uses this $S_d$ matrix as a tensor to construct an MPS. For example, we consider the following MPS
\begin{equation}
\label{eq:eq9}
\begin{split}
A^{\left[1\right]0}=v &, A^{\left[1\right]1}=v S_d, \\
A^{\left[i\right]0}=I , A^{\left[i\right]1}=S_d \ \ &\text{for} \ i=2,3,\ldots,N-1 ,\\
A^{\left[N\right]0}=v^T &, A^{\left[N\right]1}=S_d v^T.
\end{split}
\end{equation}
Here, $I$ denotes a identity matrix and $v\equiv(\begin{matrix}1&1&\cdots&1&1\\\end{matrix})$ is a row vector of dimension $d+1$ with all elements being $1$. 
When the trace for the tensor product in Eq.~\eqref{eq:eq9} is calculated,
\begin{equation}
\label{eq:eq10}
\psi_{x_1,x_2,\ldots,x_N}=v\left(S_d\right)^{\sum_{i=1}^{N} x_i}v^T
\end{equation}
is obtained. If $\sum_{i=1}^N x_i>d$, the tensor product $(S_d)^{\sum_{i=1}^N x_i}$ becomes a zero matrix and the right hand of Eq.~\eqref{eq:eq10} is $0$. Otherwise, a sub-diagonal component with the value of $1$ remains according to Eq.~\eqref{eq:eq8}, and the product becomes non-zero. Thus, the MPS described above is found to be fully feasible for the inequality constraint $\sum_{i=1}^N x_i\le d$ because Eq.~\eqref{eq:eq3} is satisfied. Figure~\ref{fig:figure2} illustrates an example under $x_1+x_2+x_3 \leq 2$. 
While the conventional method described in the paper~\cite{lopez2023tnopt} uses a nilpotent matrix at some sites as shown in Eq.~\eqref{eq:eq5}, we use the matrix at all sites, as proposed in~\cite{bachmayr2022particle}. Below, we demonstrate that the nilpotent-matrix method can be extended to linear inequality or equality constraints.

\subsection{Feasible MPS for linear inequality constraint}
\label{sec:subsec3.1}
\begin{figure}[ht]
\centering
\includegraphics[width=8cm]{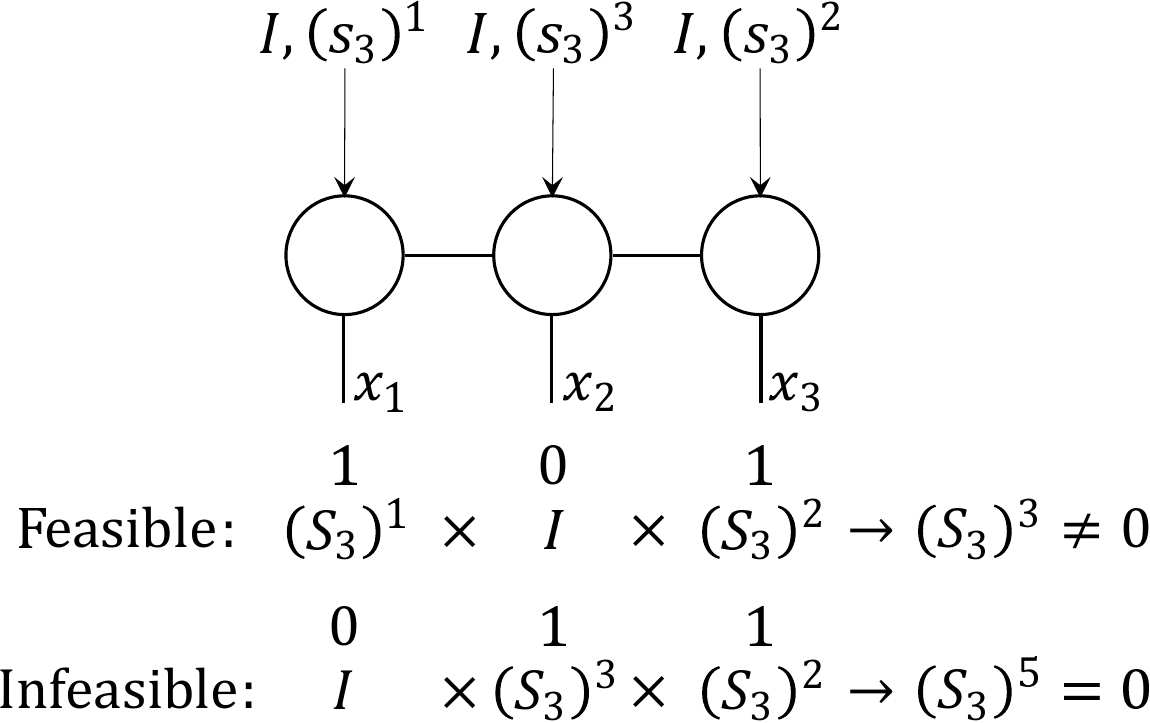}
\caption{MPS constructed by nilpotent-matrix method (positively weighted).
A special constraint $x_1+3x_2+2x_3 \leq 3$ is considered. 
A nilpotent matrix $S_3$ becomes zero when multiplied more times than upper bound $3$. 
}
\label{Nilpotent_matrix_MPS_Positive}
\label{fig:figure3}
\end{figure}

In this section, we construct a fully feasible MPS for linear inequality constraints using the nilpotent-matrix method.
\begin{equation}
\label{eq:eq11}
\sum_{i=1}^{N}{a_ix_i}\le d .
\end{equation}

First, let all coefficients $a_i$ be non-negative integers, and let $d$ be an integer greater than $0$. If $d>\sum_{i=1}^N a_i$, Eq.~\eqref{eq:eq11} becomes a trivial equation. Therefore, we focus on the case where $d\le\sum_{i=1}^N a_i$. A fully feasible MPS is defined as follows
\begin{equation}
\label{eq:eq12}
\begin{split}
A^{\left[1\right]0}=v &, A^{\left[1\right]1}=v\left(S_d\right)^{a_1}, \\
A^{\left[i\right]0}=I , A^{\left[i\right]1}=\left(S_d\right)^{a_i} &\ \ \text{for} \ i=2,3,\ldots,N-1, \\
A^{\left[N\right]0}=v^T &, A^{\left[N\right]1}=\left(S_d\right)^{a_N}v^T .
\end{split}
\end{equation}
An example under the constraint $x_1+3x_2+2x_3 \leq 3$ is shown in Fig.~\ref{fig:figure3}.

When the trace for the tensor product in Eq.~\eqref{eq:eq12} is calculated,
\begin{equation}
\label{eq:eq13}
\psi_{x_1,x_2,\ldots,x_N}=v\left(S_d\right)^{\sum_{i=1}^N{a_ix_i}}v^T
\end{equation}
is obtained. 
If $\sum_{i=1}^N{a_ix_i}>d$, the tensor product $(S_d)^{\sum_{i=1}^N{a_ix_i}}$ becomes a zero matrix and the right hand of Eq.~\eqref{eq:eq13} is $0$. Otherwise, a sub-diagonal component with the value of $1$ remains according to Eq.~\eqref{eq:eq8}, and the product becomes non-zero. That is, the above MPS is found to be fully feasible for the inequality constraint $\sum_{i=1}^N{a_ix_i}\le d$ because Eq.~\eqref{eq:eq3} is satisfied. 

\begin{figure}[ht]
\centering
\includegraphics[width=8cm]{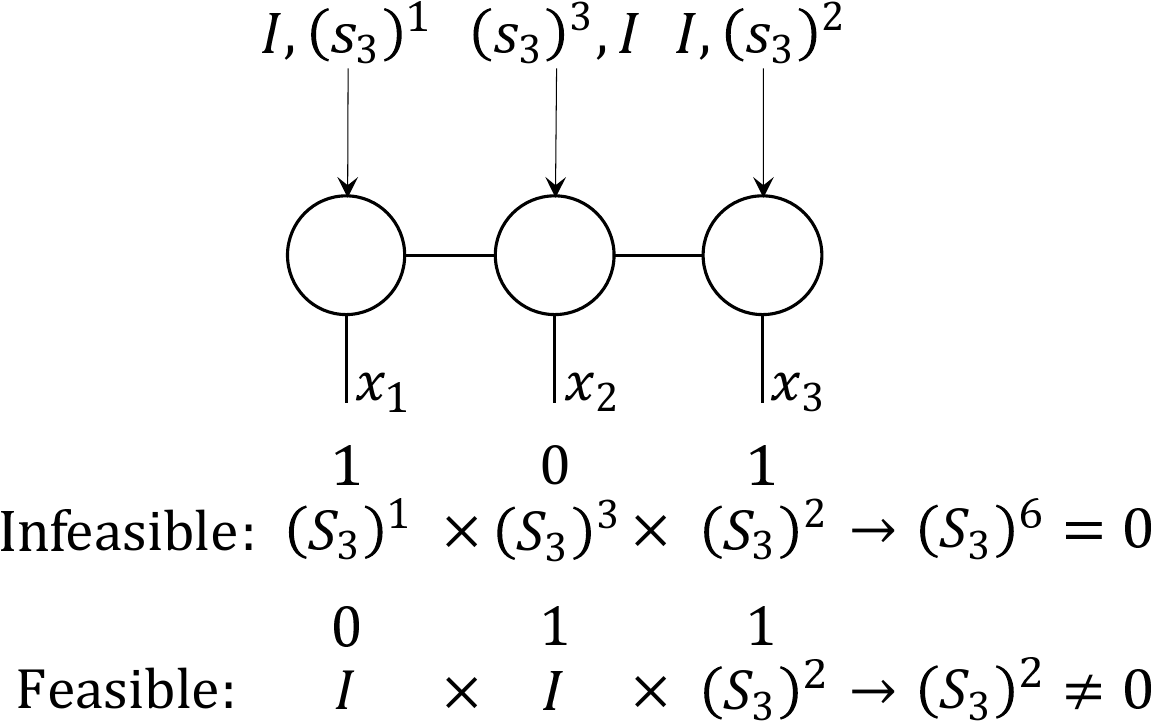}
\caption{MPS constructed by nilpotent-matrix method (arbitrarily weighted).
A special constraint $x_1-3x_2+2x_3 \leq 0$ is considered. This constraint is equivalent to $x_1+3(1-x_2)+2x_3 \leq 3$.
A nilpotent matrix $S_3$ becomes zero when multiplied more times than upper bound $3$. 
}
\label{Nilpotent_matrix_MPS_Negative}
\label{fig:figure4}
\end{figure}

Next, the constraint condition $\sum_{i=1}^N{a_ix_i}\le d$ is extended to allow negative integer coefficients. By excluding a trivial domain, we assume that $d$ is an integer such that $-\sum_{i\in\Delta_-}\left|a_i\right|\le d\le\sum_{i\in\Delta_+} a_i$. 
Here, $\Delta_+,\Delta_-$ are the index sets where the coefficient $a_i$ is non-negative and negative, respectively.
The constraint condition is equivalent to
\begin{equation}
\label{eq:eq14}
\sum_{i\in\Delta_+} a_ix_i+\sum_{i\in\Delta_-}\left|a_i\right|(1-x_i)\le d+\sum_{i\in\Delta_-}\left|a_i\right|.
\end{equation}
From Eq.~\eqref{eq:eq14}, by flipping the bit of the physical variable with a negative coefficient, the constraint condition can be converted to an equivalent condition with all non-negative coefficients. Therefore, a fully feasible MPS is defined as follows
\begin{equation}
\label{eq:eq15}
\scalebox{0.9}{$
\begin{split}
A^{\left[1\right]0}=v^\prime \left(S_{d^\prime}\right)^{|a_1|-b_1} &, A^{\left[1\right]1}=v^\prime \left(S_{d^\prime}\right)^{b_1}, \\
A^{\left[i\right]0}=\left(S_{d^\prime}\right)^{|a_i|-b_i} , A^{\left[i\right]1}=&\left(S_{d^\prime}\right)^{b_i} \ \ \text{for} \ i=2,\ldots,N-1,\\
A^{\left[N\right]0}=\left(S_{d^\prime}\right)^{|a_N|-b_N} (v^\prime &) ^T ,A^{\left[N\right]1}=\left(S_{d^\prime}\right)^{b_N}(v^\prime)^T .
\end{split}
$}
\end{equation}
Here, we let $d^\prime  \equiv d+\sum_{i\in\Delta_-}\left|a_i\right|\geq0$ and $b_i  \equiv \max{\left\{0,a_i\right\}}$,  $v^\prime \equiv(\begin{matrix}1&1&\cdots&1&1\\\end{matrix})$ is a row vector of dimension $d^\prime+1$ with all elements being $1$. 
An example under the constraint $x_1-3x_2+2x_3 \leq 0$ is shown in Fig.~\ref{fig:figure4}.

When the trace for the tensor product in Eq.~\eqref{eq:eq15} is calculated,
\begin{equation}
\label{eq:eq16}
\psi_{x_1,x_2,\ldots,x_N}=v^\prime \left(S_{d^\prime}\right)^{\sum_{i=1}^{N}{a_ix_i}+d^\prime-d}(v^\prime)^T
\end{equation}
is obtained. 
If $\sum_{i=1}^N{a_ix_i}>d$, the tensor product $(S_{d^\prime})^{\sum_{i=1}^N{a_ix_i}+d^\prime-d}$ becomes a zero matrix and the right hand of Eq.~\eqref{eq:eq16} is $0$. Otherwise, a sub-diagonal component with the value of $1$ remains according to Eq.~\eqref{eq:eq8}, and the product becomes non-zero. That is, the above MPS is found to be fully feasible for the inequality constraint $\sum_{i=1}^N{a_ix_i}\le d$ because Eq.~\eqref{eq:eq3} is satisfied. 

Finally, the case of inequality constraints that consider not only the upper bound but also the lower bound ($d_1\le d_2$) is considered.
\begin{equation}
\label{eq:eq17}
d_1\le\sum_{i=1}^{N}{a_ix_i}\le d_2 .
\end{equation}
By excluding a trivial domain, we assume that $d_1$ and $d_2$ are integers that satisfy $-\sum_{i\in\Delta_-}\left|a_i\right|\le d_1\le d_2\le\sum_{i\in\Delta_+} a_i$. In this case, a fully feasible MPS is defined as follows
\begin{equation}
\label{eq:eq18}
\scalebox{0.8}{$
\begin{split}
A^{\left[1\right]0}=v_1 \left(S_{d_2^\prime}\right)^{|a_1|-b_1} &, A^{\left[1\right]1}=v_1 \left(S_{d_2^\prime}\right)^{b_1}, \\
A^{\left[i\right]0}=\left(S_{d_2^\prime}\right)^{|a_i|-b_i} , A^{\left[i\right]1}=&\left(S_{d_2^\prime}\right)^{b_i} \ \ \text{for} \ i=2,\ldots,N-1,\\
A^{\left[N\right]0}=\left(S_{d_2^\prime}\right)^{|a_N|-b_N}(v_2)^T &, A^{\left[N\right]1}=\left(S_{d_2^\prime}\right)^{b_N}(v_2)^T .
\end{split}
$}
\end{equation}
Here, $d_2^\prime
\equiv d_2+\sum_{i \in \Delta_-} \left|a_i\right|$. $v_1 \equiv (\begin{matrix}0&\cdots&0&1\\\end{matrix})$ is a row vector of dimension $d_2^\prime+1$ with the last component being $1$ and the other components being $0$. $v_2 \equiv (\begin{matrix}1&\cdots&1&0&\cdots&0\\\end{matrix})$ is a row vector of dimension $d_2^\prime+1$ with the last $d_1^\prime \equiv d_1+\sum_{i\in\Delta_-}\left|a_i\right|$ components being $0$ and the other components being $1$.

When the trace for the tensor product in Eq.~\eqref{eq:eq18} is calculated,
\begin{equation}
\label{eq:eq19}
\psi_{x_1,x_2,\ldots,x_N}=v_1 \left(S_{d_2^\prime}\right)^{\sum_{i=1}^{N}{a_ix_i}+d_2^\prime-d_2}(v_2)^T
\end{equation}
is obtained. 
If $\sum_{i=1}^N{a_ix_i}>d_2$, the tensor product $(S_{d_2^\prime})^{\sum_{i=1}^N{a_ix_i}+d_2^\prime-d_2}$ becomes a zero matrix and the right hand of Eq.~\eqref{eq:eq19} is $0$. 
If $\sum_{i=1}^N{a_ix_i}<d_1$, the right hand is also $0$ because the last component in $(S_{d_2^\prime})^{\sum_{i=1}^N{a_ix_i}+d_2^\prime-d_2} (v_2)^T$ become $0$ due to $\sum_{i=1}^N{a_ix_i}+d_2^\prime-d_2<d_1^\prime$. 
Otherwise, under $d_1\le\sum_{i=1}^N{a_ix_i}\le d_2$, the right hand is non-zero because the last component holds the value of $1$.
Thus, the above MPS is found to be fully feasible for the inequality constraint $d_1\le\sum_{i=1}^N{a_ix_i}\le d_2$ because Eq.~\eqref{eq:eq3} is satisfied. 

\subsection{Feasible MPS for linear equality constraint}
\label{sec:subsec3.2}
The nilpotent-matrix method can be extended to linear equality constraints.
\begin{equation}
\label{eq:eq20}
\sum_{i=1}^{N}{a_ix_i}=d .
\end{equation}
Equation~\eqref{eq:eq20} is a special case of Eq.~\eqref{eq:eq17} under $d_1, d_2 \rightarrow d$. In other words, by setting $d_1, d_2 \rightarrow d$ in Eq.~\eqref{eq:eq18}, a fully feasible MPS for the constraints of Eq.~\eqref{eq:eq20} is obtained.
\begin{equation}
\label{eq:eq21}
\scalebox{0.9}{$
\begin{split}
A^{\left[1\right]0}=&\left(\begin{matrix}0&\cdots&0&1\end{matrix}\right) \left(S_{d^\prime}\right)^{|a_1|-b_1}, \\
A^{\left[1\right]1}&=\left(\begin{matrix}0&\cdots&0&1\\\end{matrix}\right) \left(S_{d^\prime}\right)^{b_1}, \\
A^{\left[i\right]0}=\left(S_{d^\prime}\right)^{|a_i|-b_i}& , A^{\left[i\right]1}=\left(S_{d^\prime}\right)^{b_i} \ \ \text{for} \ i=2,\ldots,N-1,\\
A^{\left[N\right]0}=&\left(S_{d^\prime}\right)^{|a_N|-b_N}  \left(\begin{matrix}1&0&\cdots&0\end{matrix}\right)^T, \\
A^{\left[N\right]1}&=\left(S_{d^\prime}\right)^{b_N}  \left(\begin{matrix}1&0&\cdots&0\end{matrix}\right)^T .
\end{split}
$}
\end{equation}

\section{Feasible MPS construction by shared-matrix method}
\label{sec:sec4}

In the previous section, we constructed a fully feasible MPS using nilpotent matrices to ensure the law of particle number conservation.
However, many real-world problems often involve more complex constraints than a single linear constraint. Thus, we also propose the following shared-matrix method.
This method is expected to handle complex constraints because $U(1)$ gauge symmetry is not assumed.

\subsection{Shared-matrix method}
\label{sec:subsec4.1}
\begin{figure}[ht]
\centering
\includegraphics[width=8cm]{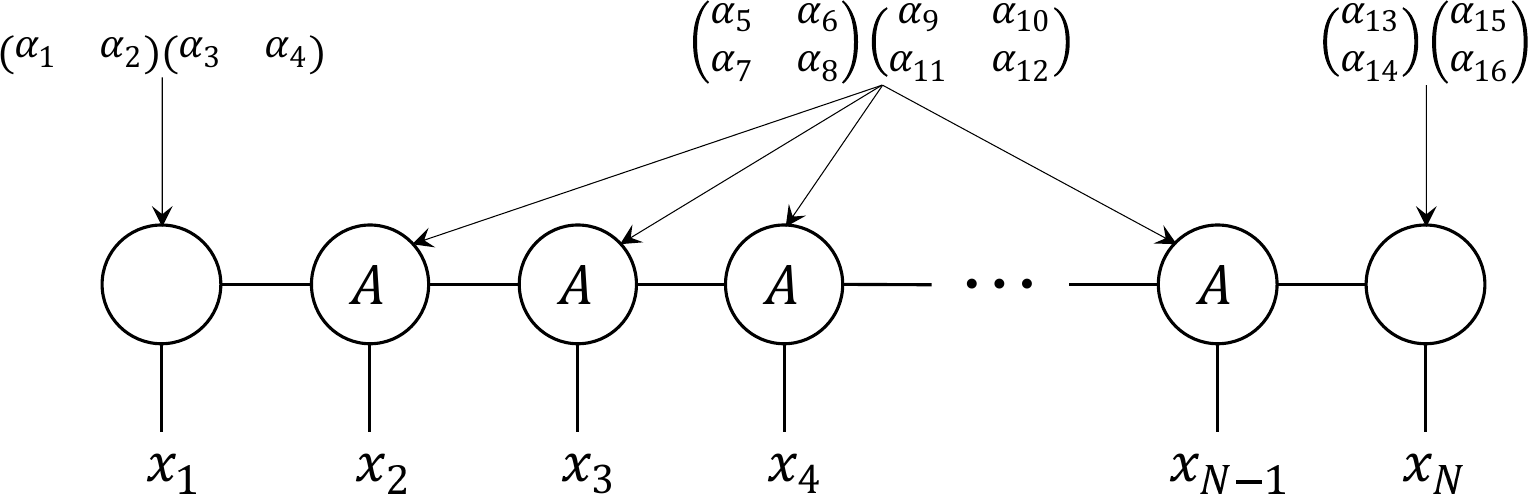}
\caption{MPS constructed by shared-matrix method. By sharing tensors on several sites, their parameters can be quickly determined so as to obtain feasible MPSs.}
\label{Shared_matrix_MPS}
\label{fig:figure5}
\end{figure}
As shown in Fig.~\ref{fig:figure5}, the shared-matrix method adopts shared matrices for the tensors of the MPS at sites except for both ends and determines the elements of each matrix inversely so that the output states become feasible solutions. That is, for an $N$-bit problem, we assume the tensors of the MPS as
\begin{equation}
\label{eq:eq22}
\scalebox{0.8}{$
\begin{split}
A^{[1]0}=\left(\begin{matrix}\alpha_1&\alpha_2\end{matrix} \right)&,A^{[1]1}=\left(\begin{matrix}\alpha_3&\alpha_4\end{matrix} \right), \\
A^{[i]0}=\left(\begin{matrix}\alpha_5&\alpha_6 \\ \alpha_7&\alpha_8\end{matrix} \right),A^{[i]1}=&\left(\begin{matrix}\alpha_9&\alpha_{10} \\ \alpha_{11}&\alpha_{12}\end{matrix} \right)  \text{for} \ i=2,\ldots,N-1, \\ 
A^{[N]0}=\left(\begin{matrix}\alpha_{13} \\ \alpha_{14}\end{matrix} \right)&,A^{[N]1}=\left(\begin{matrix}\alpha_{15} \\ \alpha_{16}\end{matrix} \right) 
\end{split}
$}
\end{equation}
and determine the real values of matrix parameters $\bm{\alpha}$. 
Hereinafter, shared matrices are represented by $A^0=A^{[i]0}, A^1=A^{[i]1} \ \text{for}\ i=2,\ldots,N-1$.
Equations for the matrix parameters are designed so that the trace of the tensor product of Eq.~\eqref{eq:eq22} becomes non-zero for feasible solutions, and $0$ for infeasible solutions. In other words, by finding the parameters to satisfy Eq.~\eqref{eq:eq3}, a fully feasible MPS can be obtained. While a $2\times2$ shared matrix is used in Eq.~\eqref{eq:eq22}, we could also adopt a larger size matrix. For example, a nilpotent matrix in Eq.~\eqref{eq:eq9} can be considered to be a $(d+1)\times(d+1)$ shared matrix. If a small size matrix is adopted, an MPS has a moderate bond dimension, which leads to an efficient calculation during imaginary time evolution. 

Generally, the matrix parameters of an MPS are difficult to be determined inversely for the target property because the number of them is intractable. However, by using a shared matrix, this number can be significantly reduced and these parameters are expected to be efficiently solved. 
The method may be applicable to problems with high periodicity and symmetry, because it is expected that feasible TNs can be described with small independent parameters.
In the following sections, we construct fully feasible MPSs for several kinds of constraints by using shared-matrix method. For simplicity, we assume that each element of the matrix takes only the values $0$ or $1$ hereinafter.

\subsection{Feasible MPS for many-to-one comparison constraint}
\label{sec:subsec4.2}
A fully feasible MPS for the many-to-one comparison constraint
\begin{equation}
\label{eq:eq23}
x_1,x_2,\ldots,x_{N-1}\le x_N
\end{equation}
($N\geq3$) is considered. This inequality constraint is often used in assignment problems such as facility location problems~\cite{farahani2010flp}.

First, we determine the type of a shared matrix to be employed. 
If $x_i = 0$ on $i$th site ($i=2,\ldots,N-1$), the values of the other bits are unaffected. 
Additionally, if at least one or more values are taken as $1$ at the bits on $i=2,\ldots,N-1$, the value of $x_N$ should be $1$. Therefore, we assume an identity matrix as the shared matrix $A^0$ and an idempotent matrix as $A^1$. Considering these assumptions, we adopt the matrices in Appendix~\ref{sec:exp_shared_matrix},
\begin{equation}
\label{eq:eq24}
A^0=I\equiv\left(\begin{matrix}1&0\\0&1\end{matrix}\right),
A^1=P\equiv\left(\begin{matrix}1&0\\0&0\end{matrix}\right)
\end{equation}
as shared matrices. As a result, the product of the tensors on the sites $i=2,\ldots,N-1$ becomes $\left(\begin{matrix}1&0\\0&0\end{matrix}\right)$ if at least one or more bits have the value of $1$ and $\left(\begin{matrix}1&0\\0&1\end{matrix}\right)$ if all bits are $0$. Thus, equations that satisfy Eq.~\eqref{eq:eq3}
\begin{equation}
\label{eq:eq25}
\begin{split}
\alpha_1 \alpha_{13}+\alpha_2 \alpha_{14}&>0, \\
\alpha_1 \alpha_{15}+\alpha_2 \alpha_{16}&>0, \\
\alpha_1 \alpha_{13}&=0, \\
\alpha_1 \alpha_{15}&>0, \\
\alpha_3 \alpha_{13}+\alpha_4 \alpha_{14}&=0, \\
\alpha_3 \alpha_{15}+\alpha_4 \alpha_{16}&>0, \\
\alpha_3 \alpha_{13}&=0, \\
\alpha_3 \alpha_{15}&>0
\end{split}
\end{equation}
are obtained.
By calculating the parameters that satisfy Eq.~\eqref{eq:eq25},
\begin{equation}
\label{eq:eq26}
\begin{split}
A^{[1]0}=\left(\begin{matrix}1&1 \end{matrix} \right)&,A^{[1]1}=\left(\begin{matrix} 1&0 \end{matrix} \right), \\
A^{[N]0}=\left(\begin{matrix}0\\1 \end{matrix} \right)&,A^{[N]1}=\left(\begin{matrix}1\\0 \end{matrix} \right)
\end{split}
\end{equation}
are obtained. By taking the product of the tensors $A^{[i]}$ in Eq.~\eqref{eq:eq24} and \eqref{eq:eq26}, a fully feasible MPS is realized.

\subsection{Feasible MPS for domain-wall encoding constraint}
\label{sec:subsec4.3}
Next, a fully feasible MPS is constructed for the domain-wall type constraint
\begin{equation}
\label{eq:eq27}
x_1\le x_2\le\cdot\cdot\cdot\le x_N
\end{equation}
($N\geq3$), which is the comparison constraint similar to Subsection~\ref{sec:subsec4.2}. The technique of designing bit variables to satisfy this constraint is called domain-wall encoding. This encoding is frequently used in quantum annealing because the number of interactions between bits is suppressed~\cite{chancellor2019dw}.

First, we determine the type of a shared matrix to be employed. 
If at least one or more values are set to $1$ at the bits on $i=2,\ldots,N-1$, the value of $x_N$ should be $1$. Additionally, if the bits on $2\le i\le N-2$ are set to $1$, the value of $x_{i+1}$ should be $1$.
Therefore, we assume an idempotent matrix as the shared matrix $A^0$, and $A^1A^0=0$.
Considering these assumptions, we adopt the matrices in Appendix~\ref{sec:exp_shared_matrix},
\begin{equation}
\label{eq:eq28}
A^0=Q\equiv\left(\begin{matrix}0&0\\1&1\end{matrix}\right),
A^1=P\equiv\left(\begin{matrix}1&0\\0&0\end{matrix}\right)
\end{equation}
as shared matrices. As a result, the product of the tensors on the sites $i=2,\ldots,N-1$ becomes $\left(\begin{matrix}0&0\\1&1\end{matrix}\right)$ if all bits are $0$ and $\left(\begin{matrix}1&0\\0&0\end{matrix}\right)$ if one domain wall exists or all bits are $1$, and a zero matrix otherwise. 
Thus, equations that satisfy Eq.~\eqref{eq:eq3}
\begin{equation}
\label{eq:eq29}
\begin{split}
\alpha_2 \alpha_{13}+\alpha_2 \alpha_{14}&>0, \\
\alpha_2 \alpha_{15}+\alpha_2 \alpha_{16}&>0, \\
\alpha_1 \alpha_{13}&=0, \\
\alpha_1 \alpha_{15}&>0, \\
\alpha_4 \alpha_{13}+\alpha_4 \alpha_{14}&=0, \\
\alpha_4 \alpha_{15}+\alpha_4 \alpha_{16}&=0, \\
\alpha_3 \alpha_{13}&=0, \\
\alpha_3 \alpha_{15}&>0
\end{split}
\end{equation}
are obtained.
By calculating the parameters that satisfy Eq.~\eqref{eq:eq29},
\begin{equation}
\label{eq:eq30}
\begin{split}
A^{[1]0}=\left(\begin{matrix}1&1 \end{matrix} \right)&,A^{[1]1}=\left(\begin{matrix} 1&0 \end{matrix} \right), \\
A^{[N]0}=\left(\begin{matrix}0\\1 \end{matrix} \right)&,A^{[N]1}=\left(\begin{matrix} 1\\0 \end{matrix} \right)
\end{split}
\end{equation}
are obtained. By taking the product of the tensors $A^{[i]}$ in Eq.~\eqref{eq:eq28} and \eqref{eq:eq30}, a fully feasible MPS is realized.

\subsection{Feasible MPS for degree-reduction constraint}
\label{sec:subsec4.4}
Next, a fully feasible MPS is constructed for the constraint
\begin{equation}
\label{eq:eq31}
\prod_{i=1}^{N-1}x_i=x_N
\end{equation}
($N\geq3$), which aims at dimension reduction. Using this reduction, it is possible to transform a higher-order function into a lower-order function~\cite{salvador2018qa}. For example, an original problem expressed in higher order binary optimization can be transformed into quadratic order binary optimization so as to be solved by quantum annealing~\cite{dattani2019hobo}.

First, we determine the type of a shared matrix to be employed. 
If $x_i = 1$ on $i$th site ($i=2,\ldots,N-1$), the values of the other bits are unaffected. 
Additionally, if at least one or more values are set to $0$ at the bits on $i=2,\ldots,N-1$, the value of $x_N$ should be $0$. Therefore, we assume an identity matrix as the shared matrix $A^1$ and an idempotent matrix as $A^0$. Considering these assumptions, we adopt the matrices in Appendix~\ref{sec:exp_shared_matrix},
\begin{equation}
\label{eq:eq32}
A^0=R\equiv\left(\begin{matrix}1&0\\1&0\end{matrix}\right),
A^1=I\equiv\left(\begin{matrix}1&0\\0&1\end{matrix}\right)
\end{equation}
as shared matrices. As a result, the product of the tensors on the sites $i=2,\ldots,N-1$ becomes $\left(\begin{matrix}1&0\\1&0\end{matrix}\right)$ if at least one or more bits have the value of $0$ and $\left(\begin{matrix}1&0\\0&1\end{matrix}\right)$ if all bits are $1$. 
Thus, equations that satisfy Eq.~\eqref{eq:eq3}
\begin{equation}
\label{eq:eq33}
\begin{split}
\alpha_1 \alpha_{13}+\alpha_2 \alpha_{13}&>0, \\
\alpha_1 \alpha_{15}+\alpha_2 \alpha_{15}&=0, \\
\alpha_1 \alpha_{13}+\alpha_2 \alpha_{14}&>0, \\
\alpha_1 \alpha_{15}+\alpha_2 \alpha_{16}&=0, \\
\alpha_3 \alpha_{13}+\alpha_4 \alpha_{13}&>0, \\
\alpha_3 \alpha_{15}+\alpha_4 \alpha_{15}&=0, \\
\alpha_3 \alpha_{13}+\alpha_4 \alpha_{14}&=0, \\
\alpha_3 \alpha_{15}+\alpha_4 \alpha_{16}&>0
\end{split}
\end{equation}
are obtained.
By calculating the parameters that satisfy Eq.~\eqref{eq:eq33},
\begin{equation}
\label{eq:eq34}
\begin{split}
A^{[1]0}=\left(\begin{matrix}1&0 \end{matrix} \right)&,A^{[1]1}=\left(\begin{matrix} 0&1 \end{matrix} \right), \\
A^{[N]0}=\left(\begin{matrix}1\\0 \end{matrix} \right)&,A^{[N]1}=\left(\begin{matrix} 0\\1 \end{matrix} \right)
\end{split}
\end{equation}
are obtained. By taking the product of the tensors $A^{[i]}$ in Eq.~\eqref{eq:eq32} and \eqref{eq:eq34}, a fully feasible MPS is realized.

\subsection{Remarks on other constraints}
The shared-matrix method can be applied to other constraints.
Firstly, it can handle the negation of Eq.~\eqref{eq:eq20}, that is $\sum_{i=1}^N a_i x_i \neq d$.
As detailed in Appendix~\ref{sec:negation}, the constraint can be encoded into only shared matrices with $2\times2$ size.
This not-equality is a well-known constraint for disjunctive programming~\cite{balas2018disjunctive}.
Usually, by utilizing Big M method~\cite{winston2004bigm}, the constraint is converted to the equivalent linear constraints $\sum_{i=1}^N a_i x_i-d\geq1-Gb$ and $\sum_{i=1}^N a_i x_i-d\leq -1+G(1-b)$ for linear programming.
Here, $G$ is a sufficiently large constant and $b$ is a binary variable.
However, our formalism can naturally encode the constraint into a simple MPS without Big M method. 
Similarly, the congruent-type constraints (i.e. $\sum_{i=1}^N a_i x_i \not\equiv 0 \bmod 2, 0 \bmod 3, \ldots$) can also be handled, as detailed in Appendix~\ref{sec:negation}.

As shown in several examples above, the shared-matrix method has a potential to derive fully feasible MPSs with a compact tensor size even for global constraints.
In addition, this method can construct such MPSs even for nonlinear constraints or not-equal constraints.

\section{Feasible MPS synthesis}
\label{sec:sec5}
Next, we explain the design of an MPS that generates states satisfying all constraints by synthesizing TNs.
For example, although the method described in Section~\ref{sec:sec3} is originally designed for a single linear constraint, this MPS synthesis allows it to handle multiple linear constraints.
Thus, when another new constraint is added to existing constraints, redesigning fully feasible TNs from scratch is not necessary.

\subsection{Feasible MPS for uncorrelated constraints}
\label{sec:subsec5.1}
\begin{figure}[ht]
\centering
\includegraphics[width=8cm]{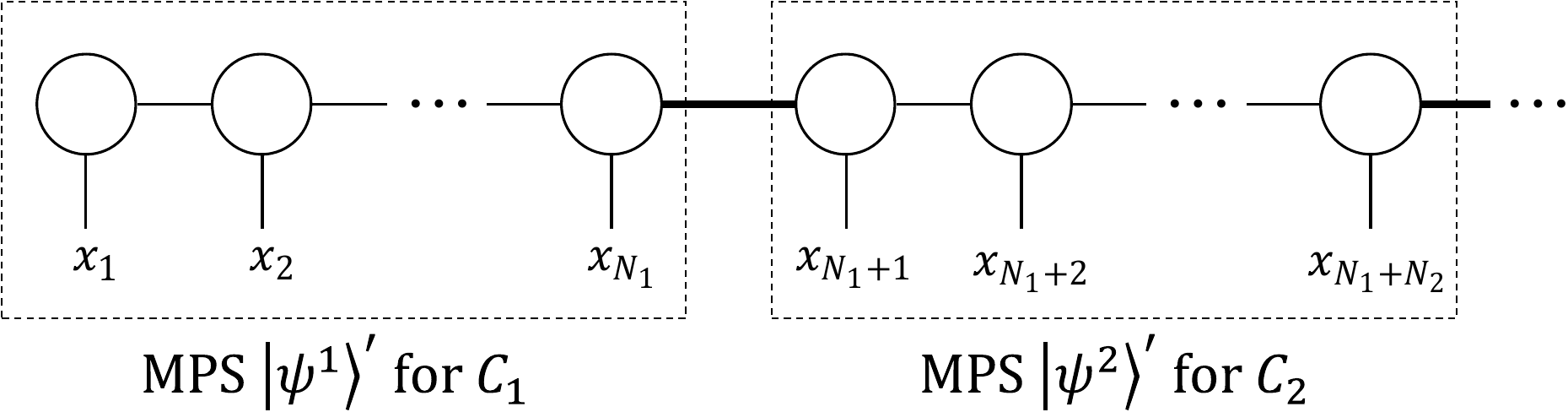}
\caption{Feasible MPS for uncorrelated constraints. A fully feasible MPS for all constraints can be constructed by connecting MPSs for each constraint in series. That is, matrix products of the tensors on the sites linked by bold bonds are performed.}
\label{Synthesized_MPS_Uncorrelated}
\label{fig:figure6}
\end{figure}

In this section, we consider a problem with multiple constraints $C_k\ (k=1,\ldots,L)$, and assume that the constraints are independent of each other. That is, there is no intersection between the index sets $D_k=\{i_1,i_2,\ldots,i_{N_k}\}$ of the physical variables appearing in each constraint $C_k$. From this assumption of independence, by appropriately arranging the order of the physical variables, the sets of physical variables appearing in each constraint $C_k$ can be enumerated in ascending order: $D_1=\{1,2,\ldots,N_1\},D_2=\{N_1+1,N_1+2,\ldots,N_1+N_2\},\ldots,D_L=\{\ldots \sum_{k=1}^L N_k\}$. In addition, the remaining $x_{1+\sum_{k=1}^L N_k},\ldots,x_N$ are the variables that do not appear in the constraints.

We assume that a fully feasible MPS has been already obtained, 
\begin{equation}
\label{eq:eq35}
\scalebox{0.8}{$
\left| \psi^k\right\rangle^\prime=\sum_{\bm{x}_{D_k}} \psi^{k}_{\bm{x}_{D_k}} \left| \bm{x}_{D_k}\right\rangle=\sum_{\bm{x}_{D_k}}\operatorname{tr}\left[\prod_{i\in D_k} \tilde{A}_k^{\left[i\right]x_i}\right] \left| \bm{x}_{D_k}\right\rangle
$}
\end{equation}
for each constraint $C_k$. Here, $\bm{x}_{D_k}=(x_{\min{D_k}},x_{\min{D_k}+1},\ldots,\ x_{\max{D_k}})$, and a prime symbol attached in the wave function is used to represent a subsystem. 
In addition, a tilde symbol attached in the tensors is used to represent a partial constraint.
A fully feasible MPS for all constraints can be constructed by connecting the above MPS for each constraint in series, as shown in Fig.~\ref{fig:figure6}.
\begin{equation}
\label{eq:eq36}
A^{\left[i\right]x_i}  =
\begin{cases}  \tilde{A}_k^{\left[i\right]x_i} & \text{for $i\in D_k$}, \\
1 & \text{for $i=\sum_{k=1}^{L}N_k+1,\ldots,N$} .
\end{cases}
\end{equation}
This is proved as follows. 
When the trace for the tensor product in Eq.~\eqref{eq:eq36} is calculated,
\begin{equation}
\label{eq:eq37}
\scalebox{0.8}{$
\begin{split}
\operatorname{tr}\left[\prod_{i=1}^N A^{\left[i\right]x_i}\ \right] &=\operatorname{tr}\left[\left(\prod_{i\in D_1} \tilde{A}_1^{\left[i\right]x_i}\right)\times\cdot\cdot\cdot\times\left(\prod_{i\in D_L} \tilde{A}_L^{\left[i\right]x_i}\right)\times1\right] \\
&=\operatorname{tr}\left[\prod_{i\in D_1} \tilde{A}_1^{\left[i\right]x_i}\ \right]\times\cdot\cdot\cdot\times \operatorname{tr}\left[\prod_{i\in D_L} \tilde{A}_L^{\left[i\right]x_i}\right] \\
&=\prod_{k=1}^{L}\psi_{\bm{x}_{D_k}}^k 
\end{split}
$}
\end{equation}
is obtained. Here, we used the property that $\prod_{i\in D_k} \tilde{A}_k^{[i]x_i}$ is not a matrix but a scalar. At the end of Eq.~\eqref{eq:eq37}, $\psi_{\bm{x}_{D_k}}^k$ is non-zero if $\bm{x}_{D_k}$ satisfies the constraint $C_k$ and $0$ otherwise. Thus, $\operatorname{tr}[\prod_{i=1}^N A^{[i]x_i}]$ satisfies Eq.~\eqref{eq:eq3}, and is proved to be a fully feasible MPS.

\subsection{Feasible MPS for correlated constraints}
\label{sec:subsec5.2}
\begin{figure}[ht]
\centering
\includegraphics[width=7cm]{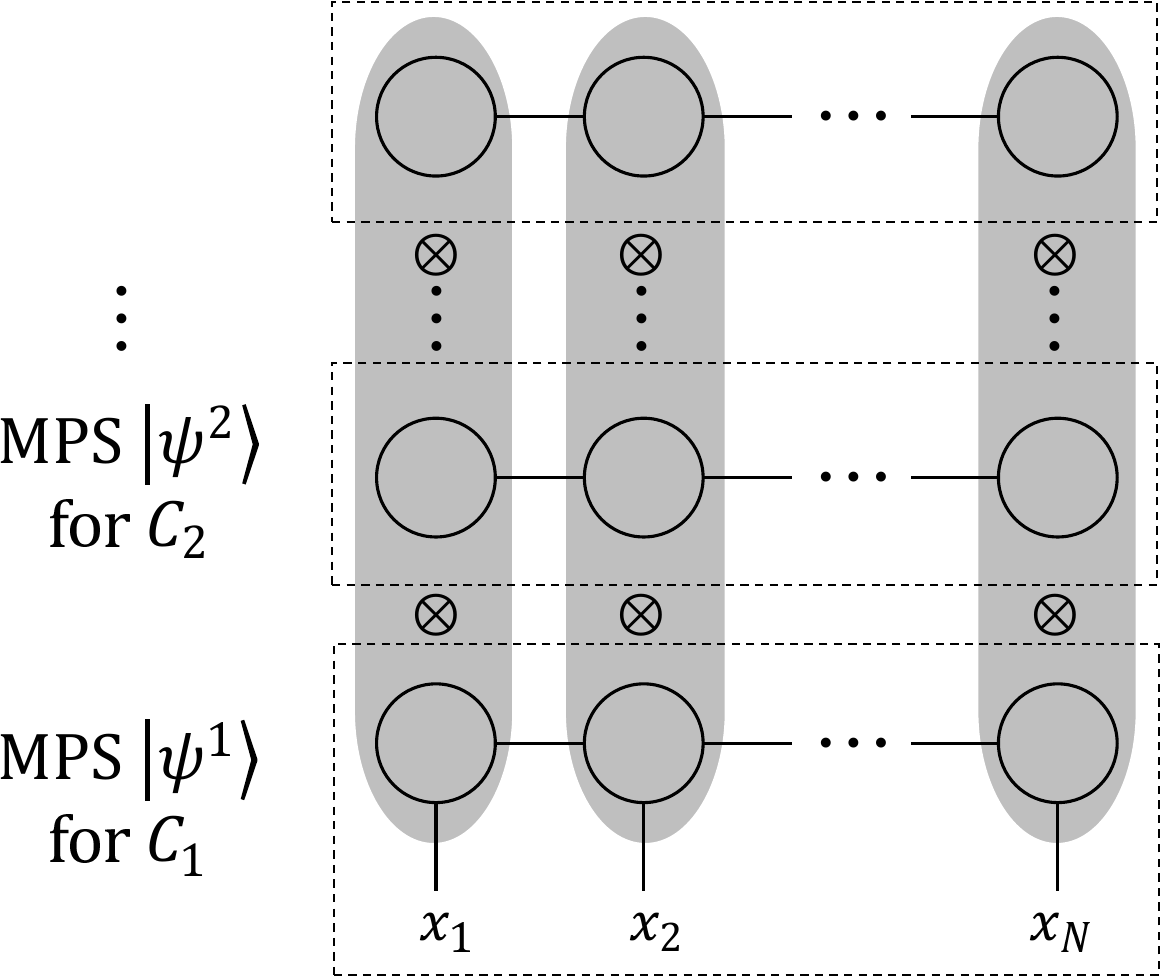}
\caption{Feasible MPS for correlated constraints. A fully feasible MPS for all constraints can be constructed by connecting MPSs for each constraint in parallel. That is, Kronecker products of the tensors inside shaded areas are performed.}
\label{Synthesized_MPS_Correlated}
\label{fig:figure7}
\end{figure}

In this section, we consider a problem with multiple constraints $C_k\ (k=1,\ldots,L)$, and assume that the constraints are not independent. We assume that a fully feasible MPS has been already obtained,
\begin{equation}
\label{eq:eq38}
\scalebox{0.8}{$
\left|\psi^k\right\rangle=\sum_{\bm{x}}{\psi^{k}_{\bm{x}}\left| x_1,\ldots,x_N\right\rangle}
=\sum_{\bm{x}}{\operatorname{tr}\left[\prod_{i=1}^{N}\tilde{A}_k^{\left[i\right]x_i}\right]\left| x_1,\ldots,x_N\right\rangle}
$}
\end{equation}
for each constraint $C_k$.
A fully feasible MPS for all constraints can be constructed by connecting the above MPS for each constraint in parallel, as shown in Fig.~\ref{fig:figure7}.
\begin{equation}
\label{eq:eq39}
A^{\left[i\right]x_i}=\tilde{A}_1^{\left[i\right]x_i}\otimes \tilde{A}_2^{\left[i\right]x_i}\cdot\cdot\cdot\otimes \tilde{A}_L^{\left[i\right]x_i} \ \ \text{for} \ i=1,2,\ldots,N .
\end{equation}
Here, $\otimes$ represents Kronecker product. 
This is proved as follows. 
When the trace for the tensor product in Eq.~\eqref{eq:eq39}, is calculated,

\begin{equation}
\label{eq:eq40}
\scalebox{0.9}{$
\begin{split}
\operatorname{tr}\left[\prod_{i=1}^N A^{\left[i\right]x_i}\ \right]&=\operatorname{tr}\left[\prod_{i=1}^N\left(\tilde{A}_1^{\left[i\right]x_i}\otimes \cdot\cdot\cdot\otimes \tilde{A}_L^{\left[i\right]x_i}\right)\ \right] \\
&=\operatorname{tr}\left[\left(\prod_{i=1}^N \tilde{A}_1^{\left[i\right]x_i}\right)\otimes\cdot\cdot\cdot\otimes\left(\prod_{i=1}^N \tilde{A}_L^{\left[i\right]x_i}\right)\right] \\
&=\operatorname{tr}\left[\prod_{i=1}^N \tilde{A}_1^{\left[i\right]x_i}\right]\times \cdot\cdot\cdot\times \operatorname{tr}\left[\prod_{i=1}^N \tilde{A}_L^{\left[i\right]x_i}\right] \\
&=\prod_{k=1}^{L}\psi_{\bm{x}}^k
\end{split}
$}
\end{equation}
is obtained. 
Here, we used the mixed-product property and spectrum property in Appendix~\ref{sec:kpd}. At the end of Eq.~\eqref{eq:eq40}, $\psi_{\bm{x}}^k$ is non-zero if $\bm{x}$ satisfies the constraint $C_k$ and $0$ otherwise. Thus, $\operatorname{tr}[\prod_{i=1}^N A^{[i]x_i}]$ satisfies Eq.~\eqref{eq:eq3}, and is proved to be a fully feasible MPS. 

Note that as shown in Eq.~\eqref{eq:eq39}, the size of the tensor increases exponentially with the number of non-independent constraints $C_k$. Therefore, for general problems involving many constraints, it is desirable to use the shared-matrix method or the technique in Subsection~\ref{sec:subsec5.1} as much as possible so as to reduce the operations of taking Kronecker product. For example, the many-to-one comparison constraint~\eqref{eq:eq23} consists of $N-1$ inequality constraints that are not independent. 
If the MPS is designed by Eq.~\eqref{eq:eq39}, the size of the tensor will be $2^{N-1}$. On the other hand, using the shared-matrix method, the MPS can be efficiently constructed by $2\times 2$ matrices as shown in Eq.~\eqref{eq:eq24}.

\subsection{Additional tensors for unconstrained variables}
\label{sec:subsec5.3}
For physical variables that do not appear in given constraints, whether the constraints are satisfied or not is independent of their values. 
Thus, one of natural choices for tensors on the corresponding site is an identity matrix. 
Their sizes are appropriately determined so that a tensor product can be well-defined.

We let the physical variable $x_i$ do not appear in the constraint. 
Additionally, we let the tensors $A_{i-1},A_{i+1}$ be already given on $i-1,i+1$th sites and their sizes be $n\times m,m \times l$, respectively.
In this case, an $m\times m$ identity matrix should be adopted as a tensor on $i$th site. 
Such formalism is ill-defined unless the column size of $A_{i-1}$ corresponds to the row size of $A_{i+1}$. However, the tensors constructed by the nilpotent-matrix method or shared-matrix method naturally satisfy this condition.

\section{Cost comparison}
\label{sec:sec6}
\begin{table*}[hbt]
  \begin{threeparttable}[h]
  \caption{Cost comparison. Here, $L$, $N$, $d$, and $k$ denote the number of constraints, the number of physical variables, the sum constant in cardinality constraint $\Sigma_i x_i = d$, and the size of shared matrices, respectively. $d^\prime_2$ is defined in Subsection~\ref{sec:subsec3.1}. $d^\prime_2$ corresponds to $d$ in the case of cardinality constraints.}
  \label{Cost_comparison}
  \label{tab:table1}
  \centering
  \begin{tabular}{cccccc} \hline
    & \cite{lopez2023tnopt} & \cite{lopez2024tnopt} & \cite{hao2022tnopt} & Nilpotent-matrix & Shared-matrix \\ \hline\hline
   Constraint & Linear & Linear & Local & Linear & Any\tnote{a} \\ \hline
   Algorithm & Backtracking & Backtracking & Adding tensor & Matrix manip. & Algebraic calc. \\ \hline
   Dimension & One & One & Higher & One & One \\  \hline
   Log size cost & $O(L \log{d})$\tnote{b} & $O(\log{N})+O(L)$ & $O(N)$ & $O(L \log{d^\prime_2})$ & $O(\log{k})$ \\ \hline
  \end{tabular}
  \begin{tablenotes}
  \item[a] \scriptsize{On condition that Eq. (3) is satisfied.}
  \item[b] \scriptsize{In the special case of multiple cardinality constraints.}
  \end{tablenotes}
  \end{threeparttable}
\end{table*}
A comparison of methods for generating fully feasible TNs is summarized in Table~\ref{tab:table1}. While the methods in~\cite{lopez2023tnopt, lopez2024tnopt} and the nilpotent-matrix method are specialized for linear constraints, the method in~\cite{hao2022tnopt} can handle various types of local constraints. The shared-matrix method can handle arbitrary constraints as long as tensor parameters to satisfy Eq.~\eqref{eq:eq3} can be appropriately derived.

Next, we compare the algorithms for obtaining fully feasible TNs. In the paper~\cite{lopez2023tnopt}, all link charges for given constraints are enumerated by backtracking. Therefore, under multiple linear constraints, this process is $\sharp$P-hard. Alternatively, the improved method~\cite{lopez2024tnopt} enumerates quantum regions, which represent the regions of a feasible charge space. In that process, backtracking is also required. The algorithm in~\cite{hao2022tnopt} checks whether each bit string satisfies the constraints or not and adds auxiliary tensors to ensure sampling feasible states. 
The proposed nilpotent-matrix method utilizes fixed nilpotent matrices. In the case of multiple constraints, Kronecker product described in Section~\ref{sec:sec5} can generate fully feasible tensors. In the shared-matrix method, the TNs are obtained by solving the equations that the parameters of the shared matrix must satisfy. 
Thus, the proposed algorithms require not backtracking but rather elementary mathematics, such as matrix manipulation and algebraic computation.

The structures of the obtained TNs are more than two-dimensional only for the method in~\cite{hao2022tnopt}, and are one-dimensional for the others. This is because, as shown in Eq.~\eqref{eq:eq6}, the method requires not only on-site tensors but also auxiliary tensors that span several sites appearing in the constraints. Thus, the proposed methods have a simpler structure and gain computational advantages by utilizing an MPS.

We discuss the cost of TNs. The maximum size of their tensors is an important factor in determining the amount of memory required to run a TN analysis.
The tensor size is exponential with respect to the number of constraints $L$ in the previous methods~\cite{lopez2023tnopt, lopez2024tnopt} and the nilpotent-matrix method, and exponential with respect to the number of physical variables $N$ in the method~\cite{hao2022tnopt}, respectively. On the other hand, the shared-matrix method has a size of $O(\log{k})$. Here, the size of the shared matrix $k$ is used.
In the several examples of the constraints in Section~\ref{sec:sec4}, $k$ is constant at $2$ regardless of the values of $L$ or $N$.
Thus, while the nilpotent-matrix method has no advantages over the previous methods with respect to a tensor size, the shared-matrix method potentially has a significant advantage.
Note that as shown in Eq.~\eqref{eq:eq39}, if Kronecker product is used for synthesizing shared matrices, the total tensor size increases exponentially with the number of producting.
The tensor size introduced above is only one form of costs. Many other costs can be considered, such as the computational cost of determining the tensor-network parameters. Definition of them will be the subject of future work.

In addition, the extensibility of the constraint conditions can be improved.
When another new constraint is added to existing constraints, the constraint is easily encoded by Kronecker product as described in Section~\ref{sec:sec5}. 
Thus, redesigning fully feasible TNs from scratch is not necessary unlike the previous methods~\cite{hao2022tnopt, lopez2023tnopt, lopez2024tnopt}.
However, note that such a MPS synthesis increases the size of tensors exponentially with respect to the number of correlated constraints.

\section{Experiment}
\label{sec:sec7}
As an application of the proposed methods, we construct a fully feasible MPS for facility location problem and perform an optimal solution search by using imaginary time evolution. Because many complex global constraints exist, such a problem is considered as a good example for the principle verification and is difficult for the previous TN-based methods to handle in principle.
In addition, the purpose of this study is to increase the number of available constraints that can be handled by TNs. Note that we do not claim any performance advantages over conventional non-TN based solvers in facility location problem.

\subsection{Fully feasible MPS for facility location problem}
\label{sec:subsec7.1}
In facility location problem~\cite{farahani2010flp}, $x_{i,j}$ is a binary variable that takes $1$ if $j$th customer is assigned to $i$th facility and $0$ otherwise, and $y_i$ is a variable that takes $1$ if $i$th facility is opened and $0$ otherwise. 
As the constraints, a group of conditions 
\begin{equation}
\label{eq:eq41}
x_{i,j}\le y_i\ \ \text{for}\ i=1,\ldots,M,j=1,\ldots,N,
\end{equation}
\begin{equation}
\label{eq:eq42}
x_{1,j}+x_{2,j}\cdot\cdot\cdot+x_{M,j}=1\ \ \text{for}\ j=1,\ldots,N
\end{equation}
is imposed. 
Here, $M$ is the total number of facility location candidates and $N$ is that of customers. 
Equation~\eqref{eq:eq41} is a many-to-one comparison constraint and Eq.~\eqref{eq:eq42} is a linear equality constraint. Therefore, a fully feasible MPS can be constructed by combining the nilpotent-matrix method and shared-matrix method.

As fully feasible MPSs, there are several ways to construct them based on the order of encoding constraint conditions. That is, the forms of the MPSs varies depending on whether Eq.~\eqref{eq:eq41} or \eqref{eq:eq42} is encoded first.  
First, we explain the former case. 
The order of the variables are rearranged as $x_{1,1},\ldots,x_{1,N},y_1,\ldots,x_{M,1},\ldots,x_{M,N},y_M$. Then, they are renamed as new physical variables $z_1,z_2,\ldots,z_{MN+M}$ for a simple representation of the MPS. 
Concretely, for the constraint~\eqref{eq:eq41}, we redefine the index set of the physical variables appearing in each constraint condition $C_i:x_{i,j}\le y_i\ \ \text{for}\ j=1,2,\ldots,N$ as $D_i$: $D_1=\{1,2,\ldots,N+1\},D_2=\{N+2,\ldots,2N+2\},\ldots,D_M=\{MN+M-N,\ldots,MN+M\}$ so as to describe the sites of the MPS in a serial order. 

The tensor of the feasible MPS for the condition $C_i$ is
\begin{equation}
\label{eq:eq43}
\scalebox{0.9}{$
\begin{split}
\tilde{A}_0^{[u_i+1]0}=\left(\begin{matrix}1&1\end{matrix} \right)&, \tilde{A}_0^{[u_i+1]1}=\left(\begin{matrix}1&0\end{matrix} \right), \\
\tilde{A}_0^{[u_i+j]0}=\left(\begin{matrix}1&0 \\ 0&1\end{matrix} \right), \tilde{A}_0^{[u_i+j]1}&=\left(\begin{matrix}1&0 \\ 0&0\end{matrix} \right) \text{for}\ j=2,\ldots,N, \\ 
\tilde{A}_0^{[u_i+N+1]0}=\left(\begin{matrix}0\\1\end{matrix} \right)&, \tilde{A}_0^{[u_i+N+1]1}=\left(\begin{matrix}1\\0\end{matrix} \right)
\end{split}
$}
\end{equation}
according to Subsection~\ref{sec:subsec4.2}. Here, $u_i \equiv (N+1)(i-1)$. Because $C_i\ (i=1,\ldots,M)$ are independent of each other, the fully feasible MPS for Eq.~\eqref{eq:eq41}
\begin{equation}
\label{eq:eq44}
\begin{split}
A_0^{[u_i+j]0} = \tilde{A}_0^{[u_i+j]0}&, A_0^{[u_i+j]1} = \tilde{A}_0^{[u_i+j]1} \\
\text{for}\  i=1,\ldots,M&,\ j=1,\ldots,N+1
\end{split}
\end{equation}
is obtained using the tensors of Eq.~\eqref{eq:eq43}.

Then, the second constraint~\eqref{eq:eq42} is encoded into this MPS. Each constraint condition $C_j^\prime:x_{1,j}+x_{2,j}\cdot\cdot\cdot+x_{M,j}=1$ is not independent from the constraint~\eqref{eq:eq41}. Therefore, it is necessary to construct a fully feasible MPS for facility location problem by taking Kronecker product as detailed in Subsection~\ref{sec:subsec5.2}. Note that because we are now using the new physical variables $\bm{z}$, each constraint condition is re-expressed as $C_j^\prime:z_{u_1+j}+z_{u_2+j}+\cdot\cdot\cdot+z_{u_M+j}=1$. 

The tensors of the fully feasible MPS for the condition $C_j^\prime$ are
\begin{equation}
\label{eq:eq45}
\scalebox{0.8}{$
\begin{split}
A_j^{[k]0}=1, A_j^{[k]1}=1  & \ \text{for}\ k=1,\ldots,j-1, \\
A_j^{[j]0}=\left(\begin{matrix}0&1\end{matrix} \right)&,A_j^{[j]1}=\left(\begin{matrix}1&0\end{matrix} \right), \\
A_j^{[u_i+j]0}=\left(\begin{matrix}1&0 \\ 0&1\end{matrix} \right),A_j^{[u_i+j]1}&=\left(\begin{matrix}0&0 \\ 1&0\end{matrix} \right) \text{for}\ i=2,\ldots,M-1, \\ 
A_j^{[u_i+k]0}=\left(\begin{matrix}1&0 \\ 0&1\end{matrix} \right)&,A_j^{[u_i+k]1}=\left(\begin{matrix}1&0 \\ 0&1\end{matrix} \right) \\
\text{for}\ i=2,\ldots,M-1, k=&1,\ldots,j-1,j+1,\ldots,N+1, \\ 
A_j^{[u_M+j]0}=\left(\begin{matrix}1\\0\end{matrix} \right)&,A_j^{[u_M+j]1}=\left(\begin{matrix}0\\1\end{matrix} \right), \\ 
A_j^{[u_M+k]0}=1, A_j^{[u_M+k]1}&=1 \ \text{for}\  k=j+1,\ldots,N+1
\end{split}
$}
\end{equation}
according to Eq.~\eqref{eq:eq21} and Subsection~\ref{sec:subsec5.3}. Substituting the tensors in Eq.~\eqref{eq:eq44} and \eqref{eq:eq45} into Eq.~\eqref{eq:eq39} gives
\begin{equation}
\label{eq:eq46}
A^{\left[i\right]z_i}=A_0^{\left[i\right]z_i}\otimes \cdot\cdot\cdot\otimes A_N^{\left[i\right]z_i} \ \ \text{for} \ i=1,\ldots,MN+M.
\end{equation}
This is one of the fully feasible MPSs for the constraints~\eqref{eq:eq41} and \eqref{eq:eq42} in facility location problem.

Next, we explain another fully feasible MPS by first encoding the constraint~\eqref{eq:eq42}. 
The order of the variables are rearranged as $x_{1,1},x_{2,1},\ldots,x_{M,1},\ldots,x_{1,N},\ldots,x_{M,N},y_1,\ldots,y_M$. Then, they are renamed as new physical variables $z_1,z_2,\ldots,z_{MN+M}$. 
Concretely, for the constraint~\eqref{eq:eq42}, we redefine the index set of the physical variables appearing in each constraint condition $C_j:\sum_{i=1}^M x_{i,j}=1$ as $D_1=\{1,2,\ldots,M\},D_2=\{M+1,\ldots,2M\},\ldots,D_N=\{MN-M+1,\ldots,MN\}$.

The tensors of the feasible MPS for the condition $C_i$ are
\begin{equation}
\label{eq:eq47}
\scalebox{0.8}{$
\begin{split}
A_0^{[w_j+1]0}=\left(\begin{matrix}0&1\end{matrix} \right)&,A_0^{[w_j+1]1}=\left(\begin{matrix}1&0\end{matrix} \right) \text{for}\ j=1,\ldots,N, \\
A_0^{[w_j+i]0}=&\left(\begin{matrix}1&0 \\ 0&1\end{matrix} \right),A_0^{[w_j+i]1}=\left(\begin{matrix}0&0 \\ 1&0\end{matrix} \right) \\
\text{for}\ i=&2,\ldots,M-1,\ j=1,\ldots,N, \\ 
A_0^{[w_j+M]0}=\left(\begin{matrix}1\\0\end{matrix} \right)&,A_0^{[w_j+M]1}=\left(\begin{matrix}0\\1\end{matrix} \right) \text{for}\ j=1,\ldots,N, \\ 
A_0^{[w_{N+1}+i]0}=1,&A_0^{[w_{N+1}+i]1}=1 \ \text{for}\ i=1,\ldots,M
\end{split}
$}
\end{equation}
according to Eq.~\eqref{eq:eq21} and Subsection~\ref{sec:subsec5.3}.  Here, $w_j \equiv M(j-1)$. 

Then, the first constraint (41) is encoded into this MPS.
Each constraint condition $C_i^\prime:x_{i,j}\leq y_i\ \ \text{for}\ j=1,\ldots,N$ is not independent from the constraint~\eqref{eq:eq42}. Therefore, it is necessary to construct a fully feasible MPS by taking Kronecker product as detailed in Subsection~\ref{sec:subsec5.2}.
Note that because we are now using the new physical variables as $\bm{z}$, each constraint condition is re-expressed as $C_i^\prime:z_{w_j+i}\leq z_{w_{N+1}+i}\ \ \text{for}\ j=1,\ldots,N$. 
The tensors of the fully feasible MPS for $C_i^\prime$ are
\begin{equation}
\label{eq:eq48}
\scalebox{0.8}{$
\begin{split}
A_i^{[k]0}=1, A_i^{[k]1}=1 \ &\text{for}\ k=1,\ldots,i-1, \\
A_i^{[i]0}=\left(\begin{matrix}1&1\end{matrix} \right)&,A_i^{[i]1}=\left(\begin{matrix}1&0\end{matrix} \right), \\
A_i^{[w_j+i]0}=\left(\begin{matrix}1&0 \\ 0&1\end{matrix} \right),A_i^{[w_j+i]1}&=\left(\begin{matrix}1&0 \\ 0&0\end{matrix} \right) \text{for}\ j=2,\ldots,N-1, \\ 
A_i^{[w_j+k]0}=\left(\begin{matrix}1&0 \\ 0&1\end{matrix} \right)&,A_i^{[w_j+k]1}=\left(\begin{matrix}1&0 \\ 0&1\end{matrix} \right) \\
\text{for}\ j=2,\ldots,N-1, k=&1,...,i-1,i+1,...,M,\\ 
A_i^{[w_N+i]0}=\left(\begin{matrix}0\\1\end{matrix} \right)&,A_i^{[w_N+i]1}=\left(\begin{matrix}1\\0\end{matrix} \right), \\ 
A_i^{[w_{N+1}+k]0}=1, A_i^{[w_{N+1}+k]1}&=1 \ \ \text{for}\  k=i+1,\ldots,M
\end{split}
$}
\end{equation}
according to Subsection~\ref{sec:subsec4.2} and \ref{sec:subsec5.3}. Substituting the tensors in Eq.~\eqref{eq:eq47} and \eqref{eq:eq48} into Eq.~\eqref{eq:eq39} gives
\begin{equation}
\label{eq:eq49}
A^{\left[i\right]z_i}=A_0^{\left[i\right]z_i}\otimes\cdot\cdot\cdot\otimes A_M^{\left[i\right]z_i} \ \ \text{for} \ i=1,\ldots,MN+M.
\end{equation}
This is another fully feasible MPS for the constraints~\eqref{eq:eq41} and \eqref{eq:eq42} in facility location problem.

We compare the two types of fully feasible MPSs. Their sizes are found to be different as shown in Eq.~\eqref{eq:eq46} and \eqref{eq:eq49}. The tensor in Eq.~\eqref{eq:eq46} is Kronecker product of $N+1$ matrices and the size is at most $2^{N+1}$, while that in Eq.~\eqref{eq:eq49} is $2^{M+1}$. This shows the amount of required memory varies depending on the order of encoding constraints. Therefore, one can choose the memory-efficient MPS based on the size difference between $M$ and $N$.
Nevertheless, in our formalism, the tensor size increases exponentially with respect to $M$ or $N$ because Kronecker products are used.

\subsection{Numerical test on fully feasible MPS construction for facility location problem}
\label{sec:subsec7.2}
\begin{figure}[ht]
\centering
\includegraphics[width=8cm]{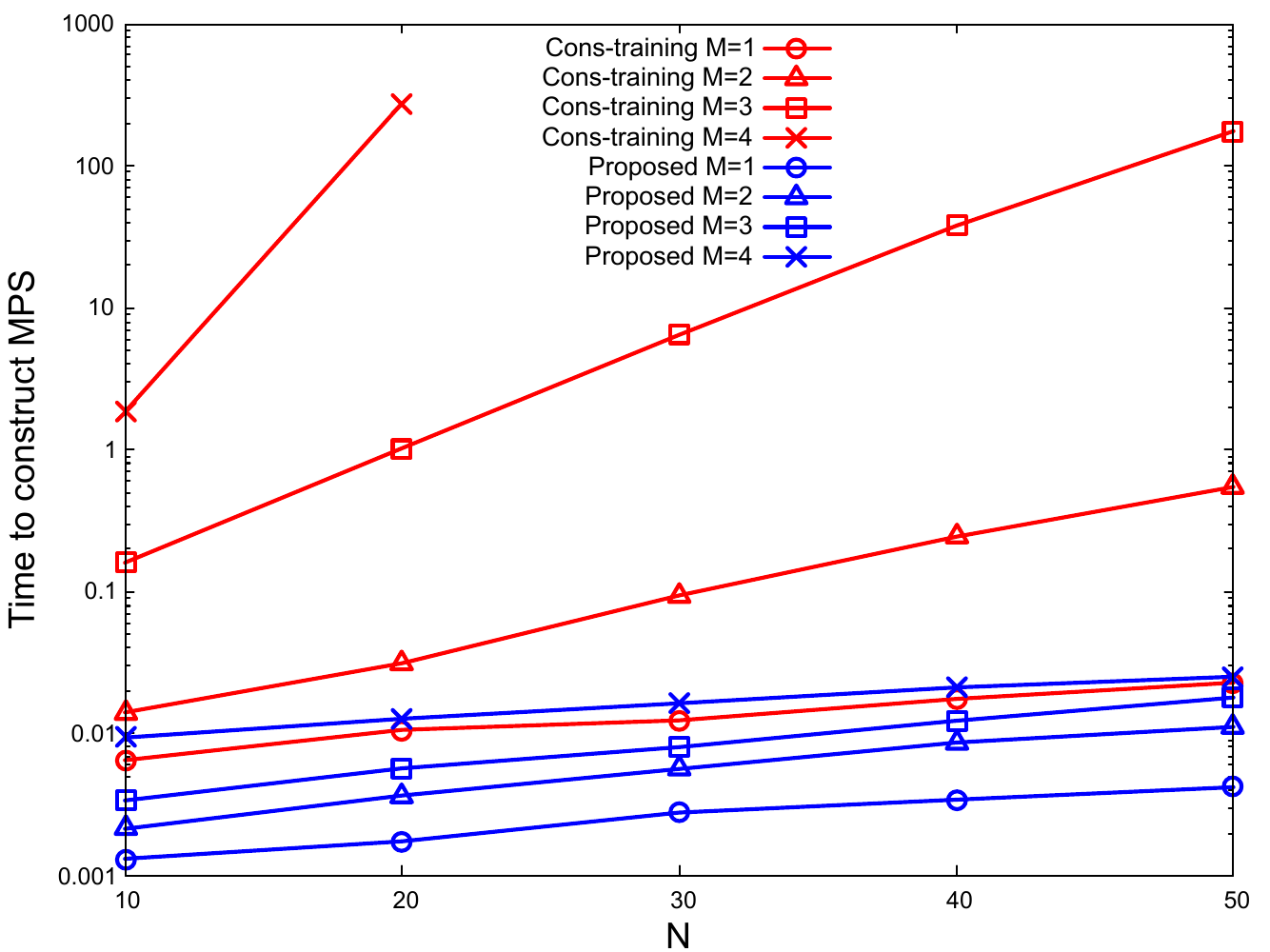}
\caption{Comparison of required times for constructing fully feasible MPSs. The unit of the vertical axis is seconds. The solid lines
shall guide the eye. In the Cons-training method, several conditions under which memory overflow occurred are not plotted.}
\label{fig:figure8}
\end{figure}

Figure~\ref{fig:figure8} shows numerical results on required times for constructing fully feasible MPSs. 
In other words, this is the time to determine all of MPS parameters for facility location constraints~\eqref{eq:eq41} and \eqref{eq:eq42}, with varying the total number of facility location candidates $M$ and total number of customers $N$.
Here, $M=1,2,3,4$ and $N=10,20,30,40,50$.
The baseline is the Cons-training method~\cite{lopez2024tnopt}, which is the modified and improved version of the method~\cite{lopez2023tnopt}. These are the methods to construct fully feasible MPSs for linear equality and inequality constraints, as mentioned in Subsection~\ref{sec:subsec2.3}.
Because Eqs.~\eqref{eq:eq41} and \eqref{eq:eq42} are linear, the Cons-training method can be applied.
This method was implemented with ConstrainTNet~\cite{lopez2024tnopt}, to obtain the fully feasible MPS.
For our method, the MPS was constructed from Eq.~\eqref{eq:eq49}, which requires a more moderate tensor size than Eq.~\eqref{eq:eq46} because $M$ is smaller than $N$. 
The simulations were performed using ITensor library~\cite{fishman2022itensor} in the Julia environment. 

The value of \texttt{maxQNs} in ITensor was set to $400$ so that MPSs could carry a sufficiently large quantum number for the Cons-training method.
After performing constructing MPSs $100$ times at each pair of $(M,N$), the averaged time was estimated and plotted with $N$ as the abscissa.
Refer to Appendix~\ref{sec:M_dependency_time} for the $M$-dependency.
For the Cons-training method, increasing $M$ up to $4$ caused calculation failures over $N=30$ due to memory overflow.
From Eqs.~\eqref{eq:eq41} and \eqref{eq:eq42}, the number of constraint equations rises to $(M+1)N$, which leads to the huge increase in time required to construct MPSs.
On the other hand, for our method, the required time was moderate. 
This may be because the shared-matrix method can encode the whole of Eq.~\eqref{eq:eq41} into $2\times2$ tensors.

In addition, the other previous method~\cite{hao2022tnopt} requires case divisions for all variables, which makes MPS construction even more intractable due to the combinatorial explosion. Specifically, in $(M,N)=(3,30)$, there are $2^{93}$ cases.
In this way, our method has efficiency superiority over conventional TN-based methods of encoding constraints into MPSs for facility location problems.

\subsection{Numerical test on optimization of facility location problem}
\label{sec:subsec7.3}
\begin{figure}[htb]
\centering
\includegraphics[width=8cm]{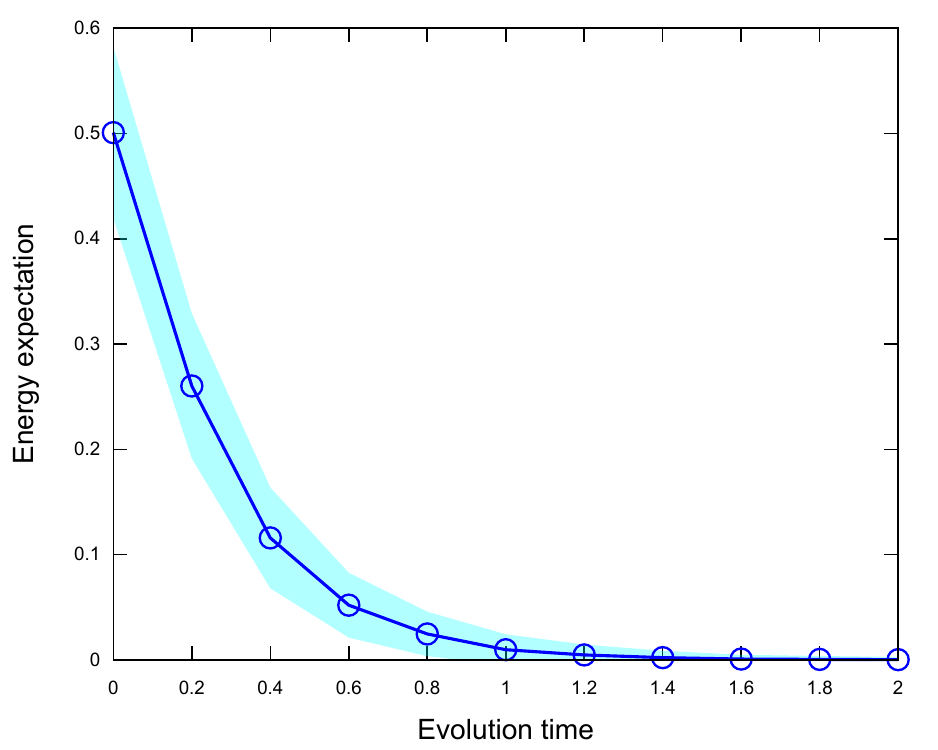}
\caption{Change of energy expectation values during imaginary time evolution. The total number of facility location candidates $M$ and total number of customers $N$ are fixed at $3$ and $40$, respectively. The solid lines shall guide the eye. A light-blue shaded area shows the standard deviations.}
\label{fig:figure9}
\end{figure}

\begin{figure}[H]
\begin{center}
\subfigure[$N=30$.]{
\includegraphics[width=7.4cm]{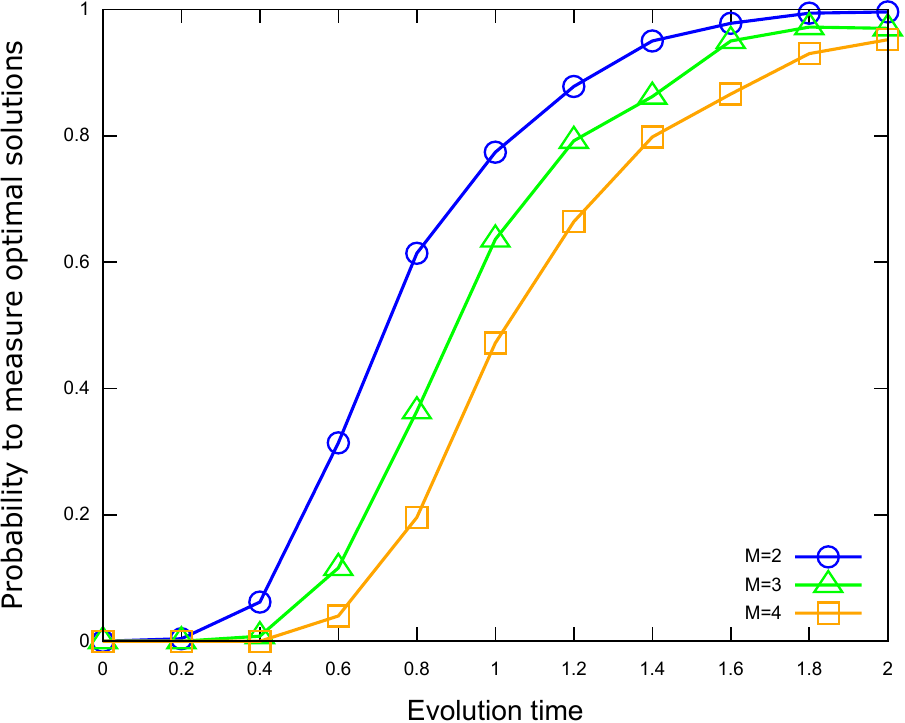}}\\
\subfigure[$N=40$.]{
\includegraphics[width=7.4cm]{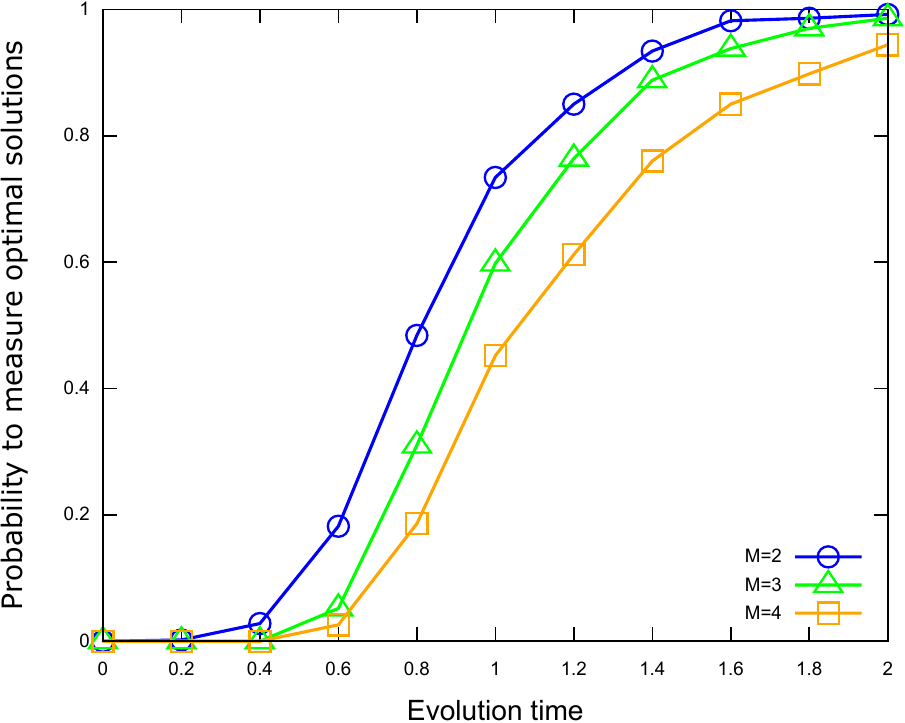}}
\subfigure[$N=50$.]{
\includegraphics[width=7.4cm]{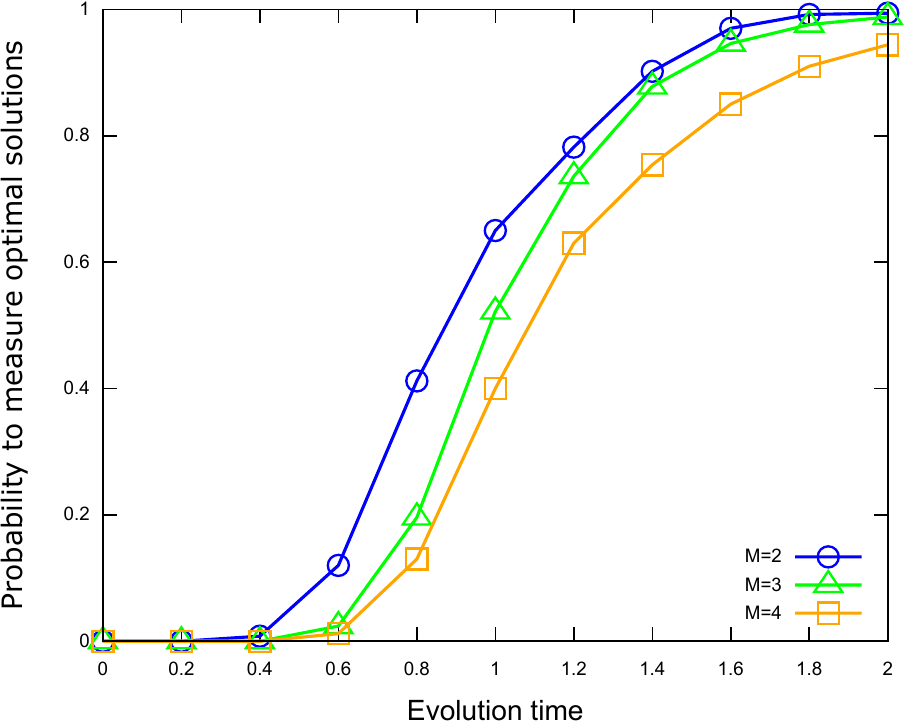}}
\caption{Probability of measuring optimal solutions for a fixed total number of customers during imaginary time evolution. (a), (b), and (c) show each plot when the total number of customers $N$ is fixed at $30$, $40$, and $50$, respectively. The solid lines shall guide the eye.} 
\label{fig:figure10}
\end{center}
\end{figure}

We performed searching for optimal solutions using the fully feasible MPS obtained in Subsection~\ref{sec:subsec7.1}. We applied imaginary time evolution to the MPS and evaluated the acquisition probabilities of the feasible solutions and optimal solutions.
The constraint conditions are given by Eq.~\eqref{eq:eq41} and \eqref{eq:eq42}, and the cost Hamiltonian to be minimized is
\begin{equation}
\label{eq:eq50}
\hat{H}=\sum_{i=1}^{M}\sum_{j=1}^{N} E_{i,j}\hat{x}_{i,j} + \sum_{i=1}^{M} F_i \hat{y}_i.
\end{equation}
For $M=2,3,4, N=30,40,50$, we determined the cost coefficients randomly and generated $500$ instances to perform the evaluation. Specifically, in Eq.~\eqref{eq:eq50}, $E_{i,j}, F_i$ were generated as an integer from the uniform distribution on $[1,3]$, and the states after imaginary time evolution were sampled.
The expected energy was calculated by substituting the state solution into Eq.~\eqref{eq:eq50}. 
We also checked whether the state was a feasible solution or an optimal solution to estimate the average probability measuring them. The results are shown in Figs.~\ref{fig:figure9} and \ref{fig:figure10}, respectively.

The optimal solutions can be easily pre-estimated by the following algorithm. First, enumerate the combinations of $\bm{y}$ by brute force. Define the index set where $y_i=1$ as $I_{y=1}$. For each $j=1,2,\ldots,N$, search for the bit string of $x_{i,j}$ with the smallest cost among $x_{i,j}\ \text{for} \ i \in I_{y=1}$. 
This gives a suboptimal solution for the fixed values of $\bm{y}$. The optimal solution is with the minimum cost among these suboptimal solutions. 
Because the total number of facility location candidates $M$ is relatively small in this experiment, the above brute force algorithm can be easily executed.

Next, we explain the details of the optimization using the proposed method. We constructed an MPS $\left| \psi\right\rangle$ using Eq.~\eqref{eq:eq49}.
For the initial state $\left| \psi\right\rangle$, we performed imaginary time evolution
\begin{equation}
\label{eq:eq51}
\left| \psi_t \right\rangle = \frac{e^{-t\hat{H}} \left| \psi \right\rangle}{\left | e^{-t\hat{H}} \left| \psi \right\rangle \right |}
\end{equation}
and sampled the final state after evolution. 

In Eq.~\eqref{eq:eq51}, the evolution time $t$ was set in increments of $0.2$ in the range $[0,2]$. 
Note that because this experiment focuses on a linear form Hamiltonian, as shown in Eq.~\eqref{eq:eq50}, it is sufficient to perform imaginary time evolution on each site locally and singular value decomposition~\cite{bridgeman2017tn} is not necessary. 
Thus, the evolution time does not require being divided into steps. In other words, $e^{-t\hat{H}} \left| \psi \right\rangle$ is calculated at once and then normalized.
If the evolution time $t$ is long, the value of the exponential function in Eq.~\eqref{eq:eq51} becomes large, leading to intractable calculation. Thus, a cutoff value was imposed: the upper limit was set to $22026$ ($\simeq e^{10}$). 
The states were sampled with ITensor's \texttt{sample} module. 

We confirmed that feasible solutions were always obtained. This is trivial because the MPS of Eq.~\eqref{eq:eq49} is feasible. 
The expectation values were min-max normalized by
\begin{equation}
\label{eq:eq52}
\epsilon_{\psi} = \frac{\braket{\hat{H}}-E_{\mathrm{min}}}{E_{\mathrm{max}}-E_{\mathrm{min}}}.
\end{equation}
Here, $E_{\text{min}}$ and $E_{\text{max}}$ denote the minimum and maximum energy in the feasible solution space, respectively. These values were calculated by the above brute force algorithm.
As shown in Fig.~\ref{fig:figure9}, the energy expectation values were confirmed to converge to zero, when $M$ and $N$ were set to $3$ and $40$, respectively.
\begin{figure}[htb]
\centering
\includegraphics[width=8cm]{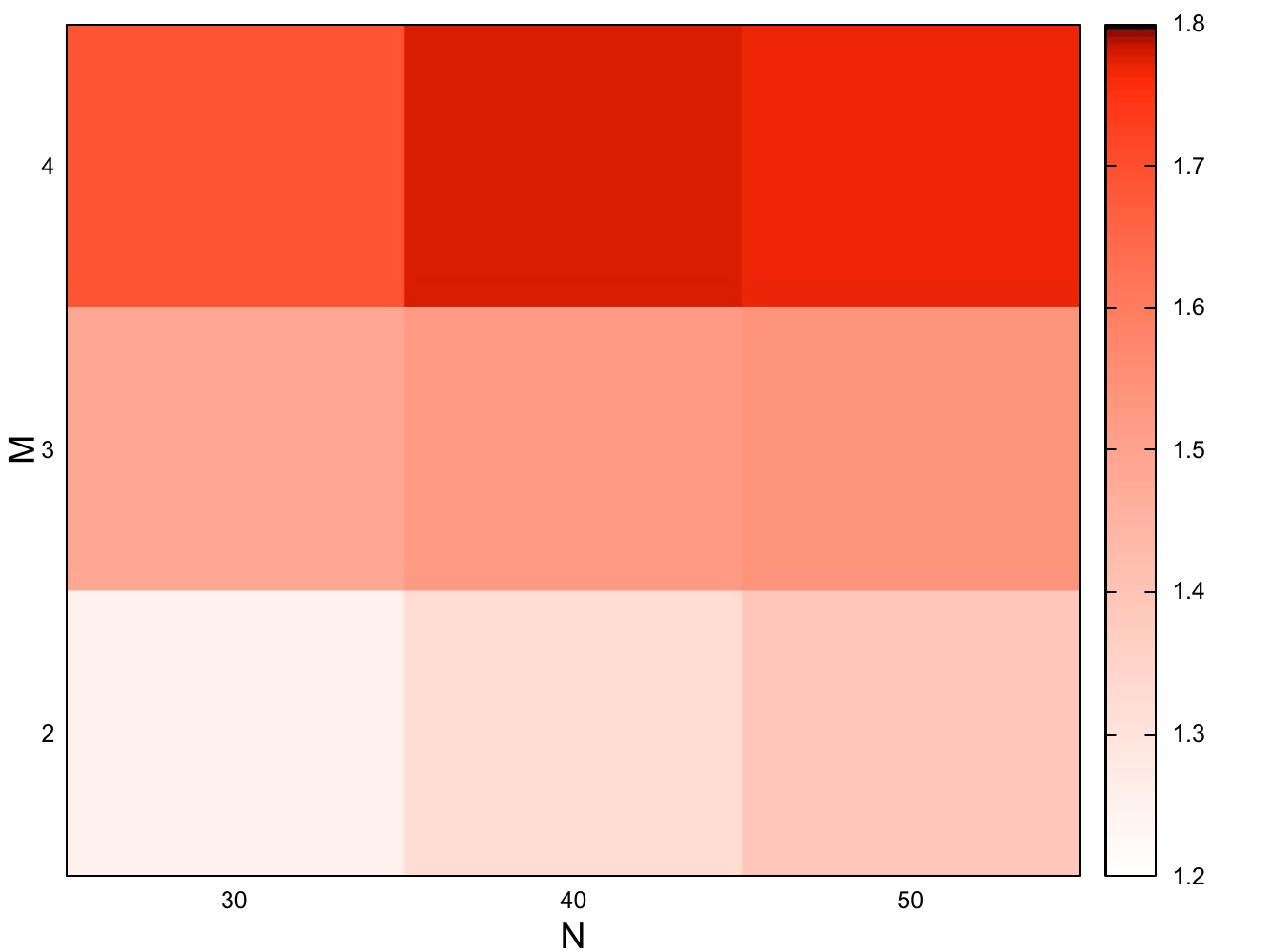}
\caption{Evolution-time heatmap for optimality. For each pair of the total number of facility location candidates $M$ and total number of customers $N$, we estimated the time such that the probability of measuring optimal solutions exceeds $0.9$. Darker colors mean longer evolution time.}
\label{fig:figure11}
\end{figure}

The probability of measuring optimal solutions was plotted, as shown in Fig.~\ref{fig:figure10}. 
The probability was found to approach $1$ along imaginary time evolution.
It was observed that the proposed method can search only for the feasible solutions and can obtain optimal solutions. 

In addition, as the total number of facility location candidates $M$ increases, the probability of measuring optimal solutions tends to decrease, as shown in Fig.~\ref{fig:figure10}.
Figure~\ref{fig:figure11} shows the evolution-time heatmap for optimality, with respect to the total number of facility location candidates $M$ and total number of customers $N$.
For each $(M,N)$, we estimated the time $t_{\rm{threshold}}$ such that the probability of measuring optimal solutions exceeds $0.9$ in the following procedure.
From Fig.~\ref{fig:figure10}, $t_{\rm{threshold}}$ was roughly determined by the intersection point with the probability $0.9$. The imaginary time evolution was then performed with varying times in increments of $0.01$ around the point, to estimate the minimum time at which the probability would be exceeded.
As shown in Fig.~\ref{fig:figure11}, $t_{\rm{threshold}}$ largely tends to increase towards the right and up.
This phenomenon is thought to be reflected by the optimization difficulty. Because the solution space expanded as the problem size increased, the convergence of optimization was considered to require longer evolution time.
Similar trends were also observed when the values of $M$ were fixed, as detailed in Appendix~\ref{sec:N_dependency_optimal}.

Finally, a comparison with the previous methods is explained.
The CBC solver~\cite{cbc2024} was used through Python library PuLP~\cite{pulp2024}.
In all instances, the optimal solutions were obtained by the CBC solver, which is reasonable because this task is a moderate-size optimization.
Thus, in the facility location problem, our method is considered to have no advantage over the conventional methods of mathematical programming.
In the future, we would like to explore problems where such advantages can be realized.

All reported experiments have been run on Sonoma 14.5 and 4.05 GHz Apple M3 CPU with 16 GB of memory.
Note that we used only a single core without parallelization.

\section{Discussion and conclusion}
\label{sec:sec8}

Recently, several methods have been proposed to solve constrained combinatorial optimization problems using TNs.
By preparing a specific TN to sample states that satisfy constraints, feasible solutions are efficiently searched for without using the penalty function methods. 
Such a TN is referred to as a feasible TN.  
These previous studies have been mainly based on profound physics, such as $U(1)$ gauge schemes and high-dimensional lattice models.
In this study, we devise to design feasible TNs using elementary mathematics without such a specific knowledge.

The nilpotent-matrix method is a technique for constructing fully feasible MPSs using nilpotent matrices. The method allows us to find MPSs specialized for a linear constraint. 
In addition, MPS synthesis in Section~\ref{sec:sec5} enables the method to handle multiple constraints.
The previous methods in~\cite{lopez2023tnopt, lopez2024tnopt} are also specialized for linear constraints and require a backtracking process to ensure $U(1)$ gauge symmetry.
On the other hand, the proposed method requires not backtracking but simple matrix manipulation.
When another new constraint is added to existing constraints, the constraint is easily encoded by Kronecker product as described in Section~\ref{sec:sec5}. 
Because redesigning fully feasible TNs from scratch is not necessary unlike the previous methods, the extensibility of the constraint conditions can be improved.

The shared-matrix method is another proposal and aims at constructing fully feasible MPSs by sharing tensor matrices across multiple sites. In this method, the matrix parameters are determined so that the trace of the MPS is non-zero for feasible solution states and $0$ otherwise. 
Because the use of shared matrices significantly reduces the number of parameters, the fully feasible MPSs can be determined by simple algebraic analysis.
Their moderate tensor size, which corresponds to that of the shared matrix, may reduce the amount of memory required to run a TN analysis.
The shared-matrix method allows us to find the fully feasible MPSs for comparison constraints (many-to-one and domain-wall encoding cases) and constraints used in degree reduction. In addition, not-equal constraints can be handled.
Our method is the first to explicitly derive fully feasible TNs for these constraints. 
The previous method in~\cite{hao2022tnopt}, which is specialized for local constraints, is difficult to apply to such global constraints.  
If the other methods in~\cite{lopez2023tnopt, lopez2024tnopt} are applied to comparison constraints, the number of linear constraints has the same order as that of physical variables $N$. 
This causes an exponential increase in the tensor size. Moreover, these methods are also difficult to apply to degree-reduction constraints or not-equal constraints because they are originally specialized for linear constraints.
Thus, the shared-matrix method has a potential to encode various types of constraints that are difficult to handle with the previous method.

In summary, the proposed methods have advantages of user-friendly design of TNs and application to a wider range of constraints.
For the principle verification, we numerically constructed a feasible TN for facility location problem, to find much faster construction than conventional methods. 
Then, by performing imaginary time evolution, feasible solutions were always obtained, ultimately leading to the optimal solution.

We discuss future issues. 
Firstly, regarding the nilpotent-matrix method, the MPS obtained by this method requires a tensor of size at most $d+\sum_{i\in\Delta_-}\left|a_i\right|+1$ for linear inequality/equality constraints with coefficients $a_i$ and constant $d$. If the value of $d$ is large, or if the absolute value of the negative coefficient is large, an increase of the tensor size may lead to intensive memory usage. Thus, more efficient encoding is required in such cases. In addition, the method of constructing fully feasible MPSs for multiple constraints by Kronecker product makes the tensor size exponentially large with respect to the number of constraints. Therefore, more efficient encoding is also required.

Although we predominantly focused on facility location problem, the application to other problems is an important issue for the proof of the versatility and superiority.
Bridging to quantum gates is also important.  In recent years, a technique for converting TNs into equivalent quantum circuits has been proposed~\cite{rudolph2023tnqc}.
By encoding our feasible MPSs into variational quantum circuits~\cite{Cerezo2021vqa}, such as Variational Quantum Eigensolver (VQE)~\cite{Peruzzo2014vqe} or QAOA~\cite{farhi2014quantum}, a new gate-based method for constrained combinatorial optimization will be realized.

\section*{Acknowledgements}
This work was partially supported by JSPS KAKENHI (Grant Number JP23H05447), the Council for Science, Technology, and Innovation (CSTI) through the Cross-ministerial Strategic Innovation Promotion Program (SIP), ``Promoting the application of advanced quantum technology platforms to social issues'' (Funding agency: QST), JST (Grant Number JPMJPF2221), and JST CREST (Grant Number JPMJCR19K4). The authors wish to express their gratitude to the World Premier International Research Center Initiative (WPI), MEXT, Japan, for their support of the Human Biology-Microbiome-Quantum Research Center (Bio2Q).

\bibliographystyle{plain}

\onecolumn
\appendix

\section{Typical examples of shared matrix}
\label{sec:exp_shared_matrix}
Here is an example of a $2\times2$ matrix used as a shared matrix. The first is an identity matrix
\begin{equation}
I \equiv \left(\begin{matrix}1&0\\0&1\\\end{matrix}\right). \nonumber
\end{equation}
It is used when the value of a physical variable does not affect whether given constraint is satisfied or not.
The second is idempotent matrices
\begin{align}
P &\equiv \left(\begin{matrix}1&0\\0&0\\\end{matrix}\right),\nonumber \\
Q &\equiv \left(\begin{matrix}0&0\\1&1\\\end{matrix}\right),\nonumber \\
R &\equiv \left(\begin{matrix}1&0\\1&0\\\end{matrix}\right). \nonumber
\end{align}
For example, an idempotent matrix is used in the case where if a certain value is taken on more than one sites, the constraint satisfaction does not depend on how many times this value is taken.
Thirdly, as a non-idempotent matrix, a nilpotent matrix
\begin{equation}
S_1 \equiv \left(\begin{matrix}0&0\\1&0\\\end{matrix}\right) \nonumber
\end{equation}
exists. For example, it is used in the case where if a certain value is taken on more than two sites, the constraint breaks.

\section{Kronecker product property}
\label{sec:kpd}
Kronecker product has several properties. The first is the mixed product property, which states for any matrices $A, B, C, D$ such that the products $AC$ and $BD$ are defined,
\begin{equation}
(A \otimes B)(C \otimes D) = AC \otimes BD \nonumber
\end{equation}
holds. Secondly, the spectral property states for any square matrices $A, B$,
\begin{equation}
\operatorname{tr}\left[A \otimes B\right] = \operatorname{tr}[A] \operatorname{tr}[B]  \nonumber
\end{equation}
is satisfied.

\section{Feasible MPS for negation of Eq.~\eqref{eq:eq20}}
\label{sec:negation}
The shared-matrix method is applied to the negation of Eq.~\eqref{eq:eq20}, that is, 
\begin{equation}
\sum_{i=1}^N a_i x_i \neq d. \nonumber
\end{equation}
Here, the coefficients $a_i$ are integer.

Firstly, the equally-weighted case $\sum_{i=1}^N x_i \neq d$ is considered. In addition, $d>N/3$ is assumed.
By using a shared matrix $A=\left(\begin{matrix}\cos{\alpha}&-\sin{\alpha}\\ \sin{\alpha}& \cos{\alpha}\end{matrix}\right)$,
\begin{equation}
\begin{split}
A^{\left[1\right]0}=\left(\begin{matrix}1&0 \end{matrix} \right)I &, A^{\left[1\right]1}=\left(\begin{matrix}1&0 \end{matrix} \right) A,  \\
A^{\left[i\right]0}=I , A^{\left[i\right]1}=A \ \ &\text{for} \ i=2,3,\ldots,N-1 ,\nonumber \\
A^{\left[N\right]0}=I\left(\begin{matrix}1\\0 \end{matrix} \right) &, A^{\left[N\right]1}=A \left(\begin{matrix}1\\0 \end{matrix} \right)
\end{split}
\end{equation}
are assumed for the feasible MPS.
Thus, 
\begin{equation}
\psi_{x_1,x_2,\ldots,x_N}=\left(A^{\sum_{i=1}^N x_i}\right)_{1,1}=\cos{\left(\alpha \sum_{i=1}^N x_i\right)} \nonumber
\end{equation}
is obtained.
To eliminate infeasible solutions, the parameter of the shared matrix is set to $\alpha=\pi/2d$. 
Because $0 \leq \alpha \sum_{i=1}^N x_i < 3\pi/2 $ is satisfied thanks to $d>N/3$, the cosine value becomes zero if and only if $\sum_{i=1}^N x_i =d$.
Thus, Eq.~\eqref{eq:eq3} is ensured and the feasible MPS is obtained.
For $d\leq N/3$, rearranging $1-x_i \to x_i$ changes the constraint into $\sum_{i=1}^N x_i=N-d$. Here, $N-d>N/3$ is satisfied and the above formulation can be applied.

Next, we consider the general case where $\sum_{i=1}a_i x_i$ is arbitrarily weighted.
If the coefficient $a_i$ is positive, as explained in Subsection~\ref{sec:subsec3.1}, $A$ is replaced by $A^{a_i}$.
On the other hand, if the coefficient $a_i$ is negative, $A$ is set to $I$ and $I$ is replaced with $A^{a_i}$.
Whereas the tensor size depends highly on $d$ for $\sum_{i=1}^N a_i x_i = d$ as shown in Section~\ref{sec:sec3}, the negation $\sum_{i=1}^N a_i x_i \neq d$ always requires only $2 \times 2$ tensors.

The above formulation can be applied to the constraint
\begin{equation}
\sum_{i=1}^N a_i x_i = \text{even}. \nonumber
\end{equation}
In this case, the parameter of the shared matrix is set to $\alpha=\pi/2$.
In addition, by changing the phase
\begin{equation}
A^{\left[1\right]0}=\left(\begin{matrix}0&1 \end{matrix} \right) , A^{\left[1\right]1}=\left(\begin{matrix}0&1 \end{matrix} \right) A, \nonumber
\end{equation}
to extract the sine instead of the cosine, the feasible MPS for $\sum_{i=1}^N a_i x_i =\text{odd}$ is obtained.
In this way, the shared-matrix method can handle the congruent-type constraints (i.e. $\sum_{i=1}^N a_i x_i \not\equiv 0 \bmod 3, 0 \bmod 4, \ldots$).

\section{$M$-dependency on time required to prepare MPSs}
\label{sec:M_dependency_time}
\begin{figure}[htb]
\centering
\includegraphics[width=8cm]{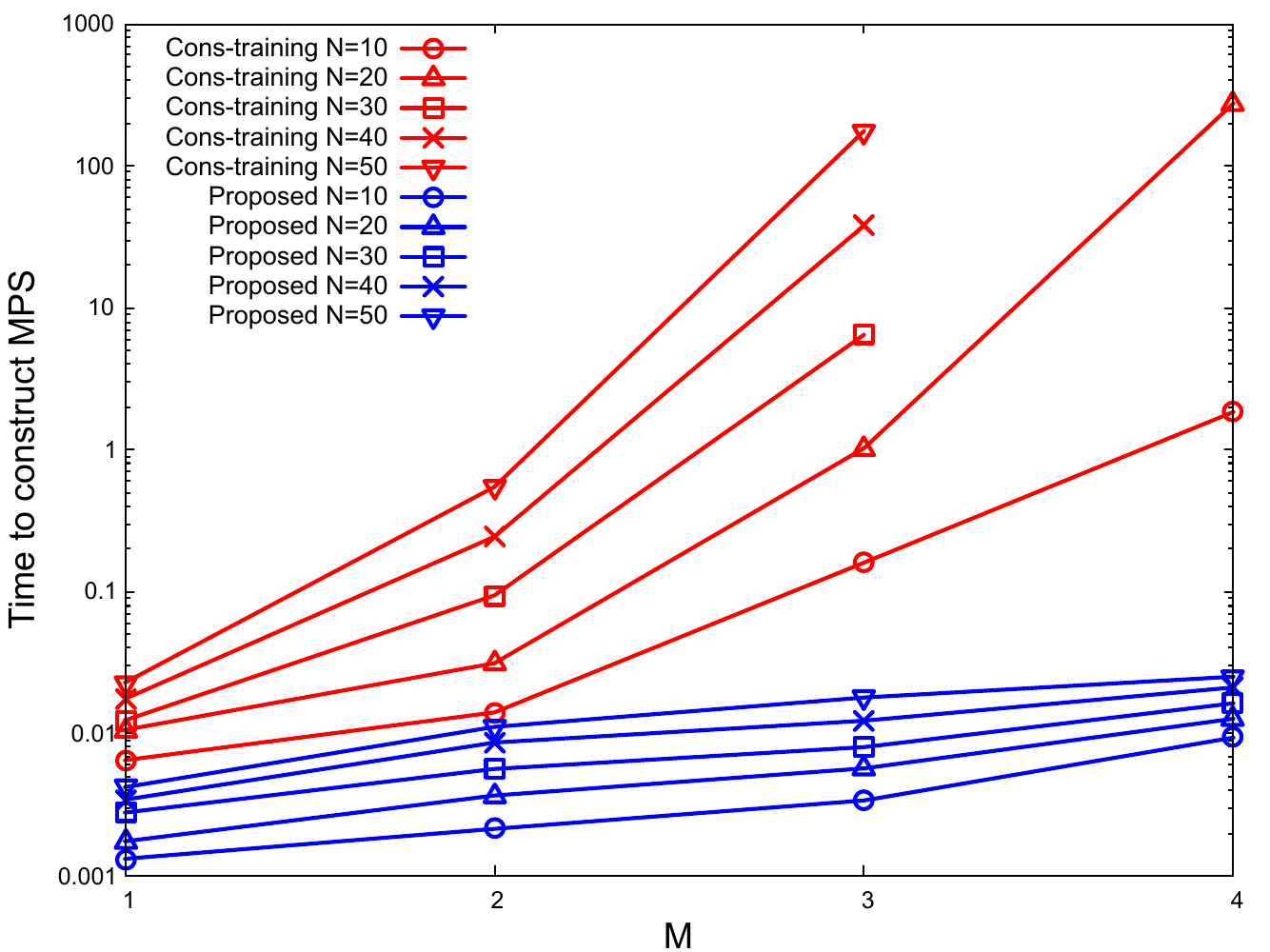}
\caption{Comparison of required times for constructing fully feasible MPSs, for a fixed total number of customers. The unit of the vertical axis is seconds. In the Cons-training method, several conditions under which memory overflow occurred are not plotted.}
\label{fig:figure12}
\end{figure}

Regarding the MPS construction for facility location problem as discussed in Subsection~\ref{sec:subsec7.2}, we explain the dependence of the total number of facility location candidates $M$ on the results.
As shown in Fig.~\ref{fig:figure12}, in the case of the Cons-training method, an increase trend larger than the exponential function can be seen.
For our method, the slope along $M$ tends to be larger than that along $N$.
Because we used Kronecker product as described in Section~\ref{sec:sec5}, the size of tensors increases exponentially  with respect to $M$. 
This would make the computational complexity sensitive to the value of $M$.
The cause of the phenomenon will be investigated in the future.
Nevertheless, our method is expected to be efficient for constructing the fully feasible MPSs of facility location problem.

\section{$N$-dependency on probability to measure optimal solutions}
\label{sec:N_dependency_optimal}

\begin{figure}[htb]
\begin{center}
\subfigure[$M=2$.]{
\includegraphics[width=7.4cm]{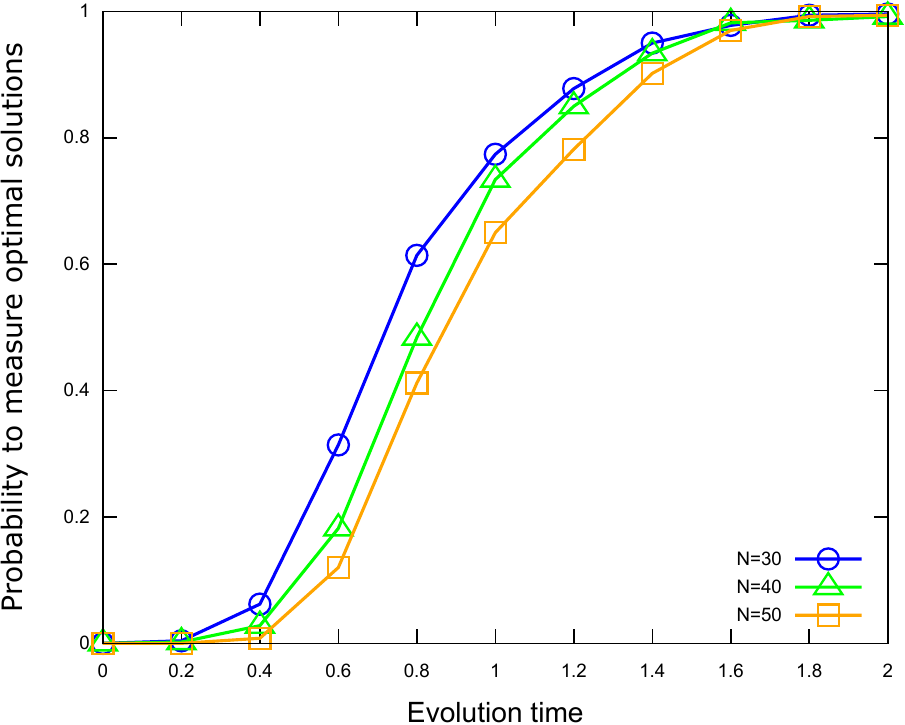}}\\
\subfigure[$M=3$.]{
\includegraphics[width=7.4cm]{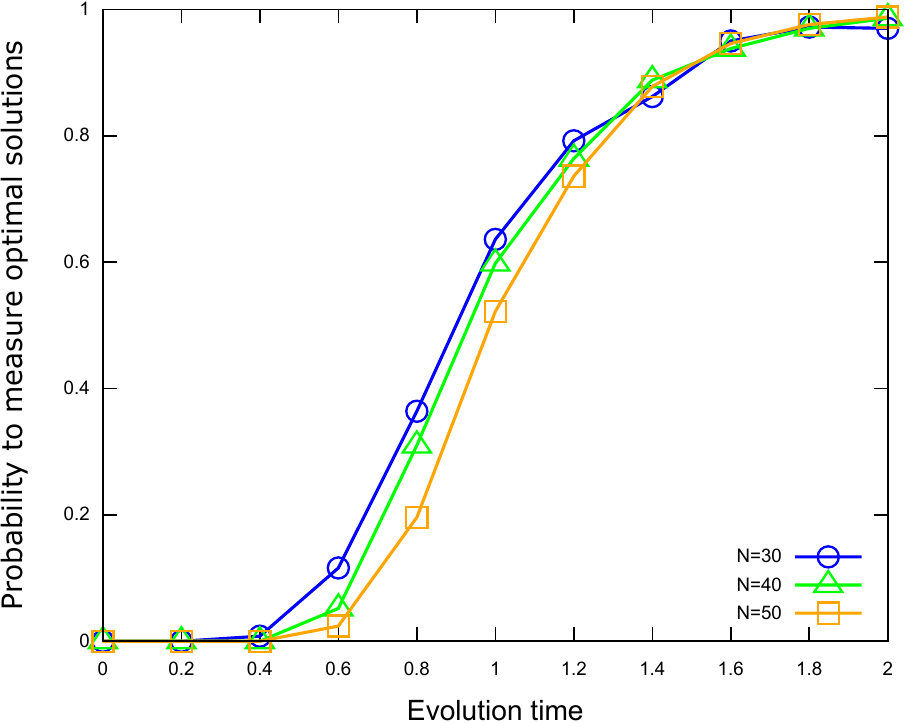}}
\subfigure[$M=4$.]{
\includegraphics[width=7.4cm]{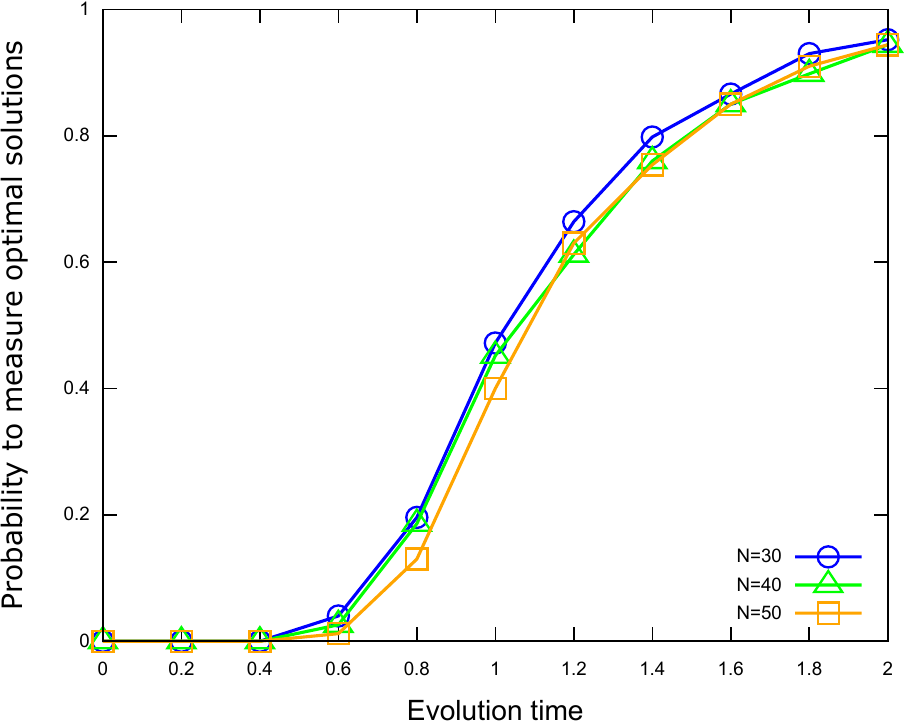}}
\caption{Probability measuring optimal solutions for a fixed total number of facility location candidates during imaginary time evolution. (a), (b), and (c) show each plot when the total number of facility location candidates $M$ is fixed at $2$, $3$, and $4$, respectively. The solid lines shall guide the eye.} 
\label{fig:figure13}
\end{center}
\end{figure}

Regarding the optimization of facility location problem as discussed in Subsection~\ref{sec:subsec7.3}, we explain the dependence of the total number of facility location candidates $M$ on the results. 
The probability of measuring optimal solutions is plotted, as shown in Fig.~\ref{fig:figure13}.
The probability was found to approach $1$ along imaginary time evolution.
Additionally, as the total number of facility location candidates $M$ increases, the probability of measuring optimal solutions tends to decrease. 
This phenomenon is thought to be reflected by the optimization difficulty, as mentioned in Subsection~\ref{sec:subsec7.3}.

\end{document}